\documentclass[%
 aip,
 amsmath,amssymb,
 reprint,%
]{revtex4-1}

\usepackage[version=3]{mhchem} 
\usepackage{physics}
\usepackage[utf8]{inputenc}
\usepackage[T1]{fontenc}
\usepackage{rotating}
\usepackage{graphicx} 
\usepackage{amsmath} 
\usepackage{etoolbox}
\usepackage{dblfnote}
\usepackage{threeparttable}
\usepackage{graphicx}    
\usepackage{float}

\newcommand{\Tp}{{\hat{T}_{\textrm{pCCD}}}}

\usepackage{multirow}

\usepackage{dcolumn}
\usepackage{bm}
\makeatother
\begin{document}

\title[Linear response pCCD-based methods]
  {Linear response pCCD-based methods: LR-pCCD and LR-pCCD+S approaches for the efficient and reliable modelling of excited state properties}

\author{Somayeh Ahmadkhani} 
\affiliation {Institute of Physics, Faculty of Physics, Astronomy and Informatics,
Nicolaus Copernicus University in Toruń, Grudziądzka 5, 87-100 Toruń, Poland}%
\author{Katharina Boguslawski}
\affiliation{%
Institute of Physics, Faculty of Physics, Astronomy and Informatics,
Nicolaus Copernicus University in Toruń, Grudziądzka 5, 87-100 Toruń, Poland}%
\author{Paweł Tecmer}
\affiliation{%
Institute of Physics, Faculty of Physics, Astronomy and Informatics,
Nicolaus Copernicus University in Toruń, Grudziądzka 5, 87-100 Toruń, Poland}%
 \email{so.ahmadkhani@gmail.com, ptecmer@fizyka.umk.pl}

\date{15 November 2024}
\doi{https://doi.org/10.1021/acs.jctc.4c01017}

\begin{abstract}
Abstract: In this work, we derive working equations for the Linear Response pair Coupled Cluster Doubles (LR-pCCD) ansatz and its extension to singles (S), LR-pCCD+S.
These methods allow us to compute electronic excitation energies and transition dipole moments based on a pCCD reference function.
We benchmark the LR-pCCD+S model against the {linear response} coupled-cluster singles and doubles method for modeling electronic spectra (excitation energies and transition dipole moments) of the BH, \ce{H2O}, \ce{H2CO}, and furan molecules.
We also analyze the effect of orbital optimization within pCCD on the resulting LR-pCCD+S transition dipole moments {and oscillator strengths} and perform a statistical error analysis. 
We show that the LR-pCCD+S method can correctly reproduce the transition dipole moments features, thus representing a reliable and cost-effective alternative to standard, more expensive electronic structure methods for modeling electronic spectra of simple molecules.
Specifically, the proposed models require only mean-field-like computational cost, while excited-state properties may approach the CCSD level of accuracy.
Moreover, we demonstrate the capability of our model to simulate electronic transitions with non-negligible contributions of double excitations and the electronic spectra of polyenes of various chain lengths, for which standard electronic structure methods perform purely. 
\end{abstract}
\maketitle
\section{Introduction}\label{sec:level1}
Electronic spectroscopy plays a pivotal role in various scientific disciplines, from chemistry to physics, biology, and astronomy. 
Electronic spectra provide a fingerprint of an investigated system.
They are crucial for understanding the behavior of electrons in atoms and molecules and the design of new complexes and materials.
Experimentally, the electronic spectrum is recorded via absorption or emission processes. 
The emission spectrum is generated when electrons transition from higher to lower energy levels, emitting photons. The absorption spectrum is produced by electrons absorbing photons to move from lower to higher energy levels.
An absorption spectrum is essentially the reverse of an emission spectrum, both provide characteristic molecular features.~\cite{electronic-spectroscopy-chem-soc-rev-1972, harris_book}

In recent years, quantum chemistry has become increasingly essential in predicting and interpreting atomic and molecular electronic spectra.~\cite{cc-for-predictions-jcp-1989, excitations-review}
Two main families of quantum chemistry methods emerged: density functional theory approximations (DFAs) and wave function theory (WFT) based methods.
While DFAs are generally more cost-effective, it is well-accepted that WFT provides more reliable results.
Specifically, the quality of DFAs' electronic spectra strongly depends on the applied approximation to the exchange--correlation functional. 
Several classes of approximate exchange--correlation functionals have been developed so far, ranging from the simplest local and semi-local, to hybrid, meta-hybrid, and double-hybrid, to more complex range-separated exchange--correlation functionals.~\cite{becke-jpc-review} 
While some of the approximations to the exchange--correlation functional provide reliable results for molecular structures, others perform well in predicting electronic excitation energies.~\cite{cam-b3lyp_spectra_liszka, pawel_saldien}
However, none of them is universal. 
Particularly challenging for DFAs are so-called multi-reference systems with a significant amount of strong electron correlation.
Such strongly-correlated electrons can be encountered, for instance, when stretching multiple bonds or in extended $\pi$-systems. 

In WFT, the exact solution to the electronic Schr\"odinger equation (in a given basis) can be obtained from the full configuration interaction (FCI) approach. 
Unfortunately, such calculations are computationally feasible only for some small model systems, comprising usually less than 20 electrons and/or orbitals, as the number of degrees of freedom scales binomially with system size. 
That technical deficiency led to the development of approximate WFT-based methods {(including the hybrid WFT/DFA methods\mbox{~\cite{mc-srDFT-formager-jcp-2013}})}, among which the coupled cluster (CC) theory~\cite{bartlett_2007} emerged as the most promising one. 
Conventional CC theory, like the coupled cluster singles and doubles (CCSD) and CCSD with perturbative triples (CCSD(T)~\cite{ccsd-parenthsis-t-cpl-1989}), can be used to accurately model large molecular systems whose electronic structures are well-represented by a single Slater determinant, that is, dominated by dynamic electron correlation effects. 
The pair Coupled-Cluster Doubles~\cite{limacher-ap1rog-jctc-2013, tamar-pccd} (pCCD) model represents a cost-effective alternative for reliable description of strongly correlated systems.~\cite{pawel-pccp-geminal-review-2022}
Combined with an efficient orbital optimization protocol,~\cite{oo-ap1rog, piotrus_mol-phys, ps2-ap1rog, ap1rog-non-variational-orbital-optimizarion-jctc} and dynamic energy correction~\cite{piotrus_pt2, ap1rog-lcc, pccd-ptx, pccd-ci, garza-pccp-wdv, pccd-tcc} pCCD-based approaches represent a versatile tool for large-scale modeling of complex systems.~\cite{pybest-paper, pccd-perspective-jpcl-2023, pccd-moco-pccp-2024, pybest-gpu-jctc-2024} 

Employing the equation of motion (EOM) formalism~\cite{rowe-eom, stanton-bartlett-eom-jcp-1993, bartlett-eom-cc-wires-2012} on top of a pCCD reference function allows us to obtain a large number of electronically excited states in one single calculation.~\cite{eom-pccd, eom-pccd-erratum, eom-pccd-lccsd}
The resulting excitation energies are size-intensive. 
An advantage of EOM-pCCD-based methods over the standard EOM-CC approaches is the low computational cost and the ability to model electronically excited states of large $\pi$-conjugated systems~\cite{pccd-delaram-rsc-adv-2023}, where the electron pair excitation energies can play a significant role.~\cite{eom-pccd, eom-pccd-lccsd} 
Unfortunately, the computation of oscillator strengths from EOM-CC methods, including the pCCD-based variants, requires the determination of both the right (as for excitation energies) and left eigenvectors of the EOM-CC equations, which essentially doubles the cost of excited-state calculations compared to the calculations of excitation energies only.~\cite{stanton-bartlett-eom-jcp-1993}
Alternatively, the electronic spectra can be obtained from the linear response (LR) formulation of CC theory (LR-CC).{\mbox{~\cite{sekino-lr-cc-ijqc-1984, koch-cc-response-jcp-1990, koch-lr-ccsd-jcp-1990, christiansen-respnse-review-ijqc-1998}}}
In that case, the excitation energies remain identical with EOM-CC,{\mbox{~\cite{watts-eom-lr-cc-review-2008, christiansen-respnse-review-ijqc-1998}}} the computation of transition dipole moments still requires left and right eigenvectors, but the resulting transition dipole moments are size-intensive.{\mbox{~\cite{koch-size-intensive-tdm-jcp-1994, christiansen-respnse-review-ijqc-1998}}}
The standard CCSD-based excitation energies can be further corrected for triple excitations using iterative\mbox{\cite{lr-cc-triplet-2}} and non-iterative techniques\mbox{~\cite{ccsdr3-jcp-1996, ccsdr3-application-water-n2-c2-cpl-1996}.}

The exceptional performance of EOM-pCCD-based methods for modeling electronic excitation energies of complex systems~\cite{eom-pccd, eom-pccd-lccsd, pccd-ee-f0-actinides} and the advantages of the LR-CC formalism motivate us to develop the LR-CC models based on a pCCD reference function (with and without orbital optimization).
In this work, we provide working equations for the LR-pCCD and LR-pCCD+S (LR-pCCD with the inclusion of singly excited states) models.
Our focus is on excited state properties, not excitation energies, which have been assessed before.~\cite{eom-pccd, eom-pccd-lccsd, pccd-ee-f0-actinides} 
We benchmark the LR-pCCD+S transition dipole moments (TDM) and oscillator strengths (OS) against the {LR-CCSD and} EOM-CCSD results for small model systems and various basis set sizes. 
Finally, we show the ability of LR-pCCD+S to model electronic spectra of all trans-polyenes of different chain lengths.


\section{Theory}\label{sec:theory}
\subsection{The pCCD ansatz}\label{sec:pccd}
One of the most interesting recent developments in CC theory~\cite{cizek_jcp_1966, cizek_paldus_1971, bartlett_2007} is the pCCD ansatz~\cite{limacher-ap1rog-jctc-2013, tamar-pccd, pawel-pccp-geminal-review-2022}, which restricts the coupled cluster wave function ansatz to only electron pair excitations and overall zero-spin, 
\begin{equation}\label{eqn:pccd}
    \ket{\textrm{pCCD}} = e^\Tp\ket{\textrm{HF}}, 
\end{equation}
where $\ket{\textrm{HF}}$ is some reference determinant (usually the Hartree--Fock determinant)
and
\begin{align}\label{eqn:tp}
    \Tp = \sum_{ai} t_{i\Bar{i}}^{a \Bar{a}} \hat{P}_a^{\dagger} \hat{P}_i.
\end{align}
In the above equation, $i, j, \ldots (a, b, \ldots)$ refer to the occupied (virtual) orbitals.  
$\hat{P}_q^{\dagger} = \hat{q}_{\uparrow}^{\dagger} \hat{q}_{\downarrow}^{\dagger}$ and $\hat{P}_p = \hat{p}_{\uparrow} \hat{p}_{\downarrow}$ are general pair creation ($\hat{P}_q^{\dagger}$) and annihilation operators ($\hat{P}_p$), respectively.
pCCD represents a computationally attractive size-extensive model with mean-field-like scaling.~\cite{limacher-ap1rog-jctc-2013}
The variational orbital optimization~\cite{oo-ap1rog, tamar-pccd, piotrus_mol-phys, ap1rog-non-variational-orbital-optimizarion-jctc, ps2-ap1rog} of the corresponding pCCD orbitals is often carried out to recover size-consistency.~\cite{tecmer2014,filip-jctc-2019} 

The pCCD state aligns with the weak formulation of the time-independent Schr\"odinger equation,
\begin{align}
    e^{-\Tp} \hat{H}_0 \ket{\textrm{pCCD}} = E_0\, e^{-\Tp} \ket{\textrm{pCCD}},
\end{align}
that is obtained by projecting the \(\bra{\textrm{HF}}\) state from the left-hand-side as \( E_0 = \bra{\textrm{HF}} \hat{H}_0 \ket{\textrm{pCCD}}\) and $\hat{H}_0$ being the molecular Hamiltonian in its normal-product form,
    \begin{align} \label{eqn:ham}
          \hat{H}_0 = & \sum_{pq} f_p^q \{ \hat{p}^{\dagger} \hat{q} \} +\frac{1}{2}\sum_{pqrs} V_{pq}^{rs} \{ \hat{p}^{\dagger} \hat{q}^{\dagger} \hat{s} \hat{r} \},
    \end{align} 
where $f_p^q$ and $V_{pq}^{rs}$ denote the Fock operator and two-electron integrals, respectively.
The pCCD amplitudes, $t_{i\Bar{i}}^{a \bar{a}}$, are then obtained by projection, 
\begin{align}\label{eqn:eqzero}
    \bra{\mu}e^{-\Tp} \hat{H}_0 \ket{\textrm{pCCD}} = 0,
\end{align}
{\mbox{where \(\bra{\mu} = \bra{\textrm{HF}} \hat{\tau}_{\mu}^{\dagger}\), and \(\hat{\tau}_{\mu}^{\dagger}\)}} refers to the de-excitation operator, here, $\hat{P}_i^\dagger \hat{P}_a$.

\subsection{LR-pCCD and LR-pCCD+S}\label{sec:lr-pccd}
The basic response equation for arbitrary operators $\hat{A}$ and $ \hat{B}$ is given by~\cite{olsen-mcscf-response-jcp-1895, koch-cc-response-jcp-1990} 
\begin{align} \label{eqn:blr}
\langle \langle \hat{A}; \hat{B}\rangle\rangle = \sum_k \bigg\{\frac{\langle 0|\hat{A}|k\rangle \langle k|\hat{B}|0\rangle}{\omega-\omega_k + i\alpha}-\frac{\langle 0|\hat{B}|k\rangle \langle k|\hat{A}|0\rangle}{\omega + \omega_k + i\alpha}\bigg\},
\end{align}
where \(\langle 0 |\) and \(|k\rangle\) are the ground and excited states, respectively, and \(\alpha\) is a real positive infinitesimal to prevent divergence. 
Defining  \(\bra{\Lambda}\) as a dual-type vector to the pCCD wave function,
\begin{align}\label{eqn:lamb}
    \langle\Lambda|    = \langle \textrm{HF} | + \sum_{\mu} \bar{t}_\mu      \langle \bar{\mu}_2 |, 
\end{align}
 satisfies the time-independent Schr\"odinger equation and the normalization condition~\cite{koch-cc-response-jcp-1990},
\begin{align}\label{eqn:norm}
    &\bra{\Lambda}\hat{H}_0\ket{\textrm{pCCD}} = E_0, \\
    &\bra{\Lambda} \ket{\textrm{pCCD}} = 1,
\end{align}
where {the index 2 in \mbox{ \(\langle \mu_2| = \langle \textrm{HF}| \tau_{\mu_2}^{\dagger}\)} refers to pair de-excitations and} \(\bar{t}_\mu\) are the Lagrange multipliers. 
Considering \(J_{\mu\nu}\) as a Jacobian matrix we will have, 
\begin{align} \label{eqn:fjac}
    \sum_{\mu} \bar{t}_{\mu} J_{\mu\nu} = \bra{\textrm{HF}} [\hat{H}_0,\hat{\tau}_{\nu}]\ket{\textrm{pCCD}}.
\end{align}
In the presence of an external potential (\(\lambda \hat{V} \)), applied to a molecular structure with \(\lambda\) intensity, Eqs.~\eqref{eqn:eqzero}, \eqref{eqn:lamb}, and \eqref{eqn:fjac} lead to

\begin{align}
    \frac{d}{d\lambda}\bra{\Lambda(\lambda)}(\hat{H}_0+\lambda\hat{V})\ket{\textrm{pCCD}(\lambda)}|_{\lambda=0} \approx \bra{\Lambda} \hat{V}\ket{\textrm{pCCD}}.
\end{align}
Using the extended Hellmann--Feynman theorem, we can incorporate \(\ket{\textrm{pCCD}}\) and \(\bra{\Lambda}\) states into the transition expectation value.
According to Eq.~(\ref{eqn:norm}), when the cluster operator is not truncated, \(\ket{\textrm{pCCD}}\) and \(\bra{\Lambda}\) states represent exact normalized states.
However, with a truncated cluster operator, \(\bra{\Lambda}\) does not act as the adjoint of \(\ket{\textrm{pCCD}}\).
In this scenario, under the assumption of a time-independent perturbation limit, the response function corresponds to the results of Refs.~\citenum{koch-cce-1990,helgaker-rgensen-1988} utilizing the Lagrangian technique approach \cite{koch-cc-response-jcp-1990}.
{In our case, the derivatives of molecular orbital coefficients with respect to the perturbation are ignored, and $\Lambda$ contains the pair excitations only for all models.}

The response function for an externally applied time-dependent potential \(\hat{B} \rightarrow \hat{V}_t\) is briefly derived in Ref.~\citenum{olsen-mcscf-response-jcp-1895} for linear and quadratic terms.
Since we are approximating the calculation to the linear part of the response function and using the coupled cluster reference wave function, the derivation of the equations is as follows
    \begin{align}\label{eqn:lreq}
        \langle\langle \hat{A};\hat{V}^{\omega}\rangle\rangle = &\sum_{n} Y_n(\omega+i\alpha) \langle \Bar{n} |\hat{A} |\textrm{pCCD}\rangle + \nonumber \\ 
        &\sum_{n} X_n(\omega+i\alpha) \langle \Lambda | [\hat{A},\hat{\tau}_{n}]|\textrm{pCCD}\rangle,
    \end{align}
where \(\hat{V}^{\omega}\) is the Fourier transformation of the time-dependent potential \(\hat{V}_t = \int \exp(-i(\omega+i \alpha)t) \hat{V}^{\omega} \, \mathrm{d}\omega\). Assuming the diagonalization of the non-symmetric Jacobian matrix, 
    \begin{align}
        J_{nm} = (U^{-1} \hat{A} U)_{nm} = \delta_{nm} \omega_n,
    \end{align}
the de-excited and excited states can be defined as \(\langle \Bar{n} | = \sum_{\nu} U^{-1}_{n \nu} \langle \Bar{\nu}|\) 
and \(\hat{\tau}_m = \sum_{\mu} \hat{\tau}_{\mu} U_{\mu m}\), respectively. 
The \(X\) and \(Y\) vectors in Eq.~(\ref{eqn:lreq}) are then expressed in terms of the Jacobian matrix and excitation energies ($\omega$),
    \begin{align}
        X_{\mu}(\omega) & = \sum_{\nu} (\boldsymbol{J}+(\omega + i \alpha) \boldsymbol{I})_{\mu\nu}^{-1}\xi_{\nu},
    \end{align}
and
    \begin{align}
        Y_{\mu}(\omega) =  \sum_{\nu} & \left(  \langle\Lambda|[\hat{V}^{\omega},\hat{\tau}_{\nu}]|\textrm{pCCD}\rangle + \sum_{\gamma} F_{\nu\gamma} X_{\nu}(\omega + i\alpha)\right)\cross \nonumber \\ 
        &(\boldsymbol{J}+(\omega + i \alpha)\boldsymbol{I})_{\mu\nu}^{-1}.
    \end{align}
Explicit forms of LR-pCCD and LR-pCCD+S vectors and matrices are collected in Table~\ref{tab:table1}. 
More details related to the derivation of these vectors and matrices can be found in the SI. 
 \begin{table*}[ht!]
   \caption{Response vectors and matrices for the LR-pCCD and LR-pCCD\(+\)S equations.
   Indices n=1 and n=2 refer to single and pair excitations (de-excitation) with respect to the \(\tau_{\mu n}\) (\(\tau^{\dagger}_{\nu n}\)) operator. {For more details see Ref.\mbox{~\citenum{koch-cc-response-jcp-1990}}}.}\label{tab:error}
   \begin{tabular}{l l l| l l}
   \\
      Name    & \hspace{0.3cm}LR-pCCD & & \hspace{0.3cm}LR-pCCD\(+\)S & \\
  \hline
    &&&& \\
     $ \, \bar{A}$ \hspace{0.4cm} = \,  & $e^{-\Tp} \,\hat{A} \,e^{\Tp}$ & & \hspace{0.2cm}  $e^{-\Tp} \,\hat{A} \,e^{\Tp}$ & \\
     &&&&\\
     $\eta^{\hat{A}}_{\nu}$ \, \, = \,  & $ \langle \textrm{HF}|[\hat{A},\hat{\tau}_{\nu_2}]|\textrm{pCCD}\rangle$ & 
                                        &\hspace{0.2cm} $ \langle \textrm{HF}|[\hat{A},\hat{\tau}_{\nu_2}]|\textrm{pCCD}\rangle $, & $ \langle \textrm{HF}|[\hat{A},\hat{\tau}_{\nu_1}]|\textrm{pCCD}\rangle $ \\
     &&&&\\
     $\xi^{\hat{A}}_{\mu}$ \, \, = \,  & $ \langle \bar{\mu}_2|\bar{A}|\textrm{pCCD}\rangle$ &
                                        &\hspace{0.2cm} $ \langle \bar{\mu}_1|\bar{A}|\textrm{pCCD}\rangle$, & $ \langle \bar{\mu}_2|\bar{A}|\textrm{pCCD}\rangle $\\
     &&&&\\
     $F_{\mu \nu}$ \, = \,              & $\langle\Lambda|[[\hat{H}_0,\hat{\tau}_{\nu_2}],\hat{\tau}_{\mu_2}]|\textrm{pCCD}\rangle$ &
                                        &\hspace{0.2cm} $\langle\Lambda|[[\hat{H}_0,\hat{\tau}_{\nu_1}],\hat{\tau}_{\mu_1}]|\textrm{pCCD}\rangle$, 
                                        & $\langle\Lambda|[[\hat{H}_0,\hat{\tau}_{\nu_1}],\hat{\tau}_{\mu_2}]|\textrm{pCCD}\rangle$ \\ 
                                        & &
                                        &\hspace{0.2cm} $\langle\Lambda|[[\hat{H}_0,\hat{\tau}_{\nu_2}],\hat{\tau}_{\mu_1}]|\textrm{pCCD}\rangle$, 
                                        & $\langle\Lambda|[[\hat{H}_0,\hat{\tau}_{\nu_2}],\hat{\tau}_{\mu_2}]|\textrm{pCCD}\rangle$ \\
     &&&&\\
     $J_{\mu \nu}$ \, = \,              & $\langle \textrm{HF}|[[\hat{H}_0,\hat{\tau}_{\nu_2}],\hat{\tau}_{\mu_2}]|\textrm{pCCD}\rangle$ & 
                                        &\hspace{0.2cm} $\langle \textrm{HF}|[[\hat{H}_0,\hat{\tau}_{\nu_1}],\hat{\tau}_{\mu_1}]|\textrm{pCCD}\rangle$, 
                                        & $\langle \textrm{HF}|[[\hat{H}_0,\hat{\tau}_{\nu_1}],\hat{\tau}_{\mu_2}]|\textrm{pCCD}\rangle$\\
                                        & &
                                        &\hspace{0.2cm} $\langle \textrm{HF}|[[\hat{H}_0,\hat{\tau}_{\nu_2}],\hat{\tau}_{\mu_1}]|\textrm{pCCD}\rangle$, 
                                        & $\langle \textrm{HF}|[[\hat{H}_0,\hat{\tau}_{\nu_2}],\hat{\tau}_{\mu_2}]|\textrm{pCCD}\rangle$\\ 
    &&&&\\
    \hline
   \end{tabular}
   \label{tab:table1}
 \end{table*}

\subsection{Excitation energies\label{sec:ex_e}} 
Opposite to the EOM-CC approach, we don’t need to diagonalize the effective Hamiltonian to obtain excitation energies using the LR-CC formalism. {We should emphasize that the excitation energies and excited states are extracted from the diagonalized Jacobian matrix.
It links the eigenvalues of the Jacobian with the underlying excited-state energies and associated eigenvectors with the excited-state wavefunctions.} To clarify this further, we can rewrite the LR function from Eq.~\eqref{eqn:blr} as
    \begin{align}
        \langle\langle \hat{A}; \hat{B}\rangle\rangle = \boldsymbol{A}^{[1]T}(\boldsymbol{E}^{[2]}-\omega \boldsymbol{S}^{[2]})\boldsymbol{B}^{[1]},
    \end{align}
where $\boldsymbol{A}^{[1]T}$ is the transposed one-dimensional array of $\hat{A}$ components $(\langle 0|\hat{A}|n\rangle, \langle n|\hat{A}|0\rangle)$, $\boldsymbol{B}^{[1]} = (\langle n|\hat{B}|0\rangle, \langle 0|\hat{B}|n\rangle)$, and the term in parenthesis provides the excitation energies of the system.
By considering this concept, we can rewrite the LR-CC function as follows
    \begin{align} \label{eqn:flin}
        \langle\langle \hat{A}; \hat{V}^{\omega}\rangle\rangle = \eta^{\hat{A}} (\boldsymbol{E}^{[2]}-\omega \boldsymbol{S}^{[2]})g^{\hat{V}^{\omega}},
    \end{align}
where $\eta_i^{\hat{A}} = (\langle 0|[\bar{{A}},\hat{\tau}_i]|0\rangle, \langle 0|[\Bar{A},\hat{\tau}_i^{\dagger}]|0\rangle $ and $g_i^{\hat{V}^{\omega T}} = (\langle 0|[\Bar{V}^{\omega},\hat{\tau}_i]|0\rangle, \langle 0|[\Bar{V}^{\omega},\hat{\tau}_i^{\dagger}]|0\rangle)$.
As a result, we obtain the excitation energies of the system by diagonalizing $(\boldsymbol{E}^{[2]}-\omega \boldsymbol{S}^{[2]})$, where 
    \begin{align}
          \boldsymbol{E}^{[2]} = E_{\mu,\nu} = \begin{pmatrix} \langle 0|[[\hat{\tau}_{\mu}, \bar{{H}}_0], \hat{\tau}_{\nu}^{\dagger}]|0\rangle & \langle 0|[[\hat{\tau}_{\mu}, \bar{{H}}_0],\hat{\tau}_{\nu}]|0\rangle \\ \langle 0|[[\hat{\tau}_{\mu}^{\dagger},\bar{{H}}_0],\hat{\tau}_{\nu}]| 0\rangle & \langle 0|[[\hat{\tau}_{\mu}^{\dagger}, \bar{{H}}_0], \hat{\tau}_{\nu}^{\dagger}]|0\rangle \\ \end{pmatrix},
    \end{align}
and
    \begin{align}
        \boldsymbol{S}^{[2]} = S_{\mu,\nu} = \begin{pmatrix} \langle 0|[\hat{\tau}_{\mu}, \hat{\tau}_{\nu}^{\dagger}]|0\rangle & \langle 0|[\hat{\tau}_{\mu}, \hat{\tau}_{\nu}]|0\rangle \\ -\langle 0|[\hat{\tau}_{\mu}^{\dagger},\hat{\tau}_{\nu}^{\dagger}]|0\rangle & -\langle 0|[\hat{\tau}_{\mu}^{\dagger}, \hat{\tau}_{\nu}]|0\rangle\\ \end{pmatrix}.
    \end{align}
The diagonal terms of the $\boldsymbol{S}$ matrix contain single excitations and de-excitations. 
The off-diagonal terms of $\boldsymbol{S}$ accommodate only pair excitations and de-excitations and are usually small. 

Let's now consider the molecular Hamiltonian in its normal-product form $\hat{H}_0$
instead of the $\hat{A}$ operator and represent the $\hat{B}$ operator by any external potential operator. 
Now, Eq.~(\ref{eqn:flin}) can be simplified such that $\boldsymbol{E}$ will refer to the Jacobian matrix, and $\boldsymbol{S}$ would become the identity matrix (see also Eq.~(4.36) of Ref.~\citenum{christiansen-respnse-review-ijqc-1998}).
Specifically for the pair and single excitation operators and effective Hamiltonian, the following relations take place, 
    \begin{align}
        \hat{\tau}_{\nu_2} & = \hat{P}_b^{\dagger} \hat{P}_j, \hspace{1cm} \hat{\tau}_{\nu_1} = \hat{b}^{\dagger} \hat{j}, \nonumber\\ 
        \bar{{H}}_0 & = e^{-\Tp} \hat{H}_0 e^{\Tp} \nonumber\\
        & = \hat{H}_0 +[\hat{H}_0,\Tp] +... . 
    \end{align}
Thus, we can investigate the excitation energies using all standard CC flavors, including pCCD models, by diagonalizing the Jacobian matrix,
    \begin{align}
        (\hat{J}-\omega \boldsymbol{I})\boldsymbol{X} = 0.
    \end{align}
The explicit Jacobian matrix elements derived for the pCCD and pCCD+S models are provided in the last row of Table~\ref{tab:table1}.
To diagonalize such a Jacobian matrix, we can use the non-symmetric Davidson method \cite{cave-ee-butadiene-jcp-1987, pccd-ci}.
We should stress here that EOM-pCCD and LR-pCCD trivially have the same set of eigenvalues (excitation energies) as the diagonalization problems are equivalent.
However, EOM-pCCD+S and LR-pCCD+S slightly differ in their excitation spectra as we only consider the Jacobian in the latter, while the former contains a non-zero first column in the effective Hamiltonian to be diagonalized.\mbox{\cite{eom-pccd,eom-pccd-erratum}}
These differences in excitation energies are minor (a few \mbox{$\mu E_h$}) and orders of magnitude smaller than the accuracy of excitation energies.

\subsection{Transition matrix elements}
To study the excited state's properties, we need to align the excitation properties with those of the ground state.
For that purpose, we use transition matrices that transfer specific ground state operator effects to the excited states.
We need the excitation energies and vectors derived in section~\ref{sec:ex_e} to achieve that. 

The linear response function of Eq.~\eqref{eqn:blr} exhibits poles at energies \(\pm \omega_k\).
By evaluating the residues of these poles, we can straightforwardly obtain transition matrix elements.
That involves approximating the frequencies of the linear response function in Eq. (\ref{eqn:lreq}) by the pole values, leading to the following relation,
    \begin{align}
         &\langle 0|\hat{A}|k\rangle \langle k|\hat{B}|0\rangle = 
         \lim_{\omega\rightarrow\omega_k} (\omega-\omega_k) \langle\langle \hat{A};\hat{B}\rangle\rangle = \nonumber \\ 
         &   -\sum_n \bigg\{\langle \Bar{n} |\hat{A}|\textrm{pCCD}\rangle (\omega-\omega_n)^{-1} F_{\nu n}  +
          \nonumber \\ 
         &\langle \Lambda | [\hat{A},\hat{\tau}_n]|\textrm{pCCD}\rangle - X_{\nu}^{\hat{A}}(-\omega_n) F_{\nu n} \bigg\} \langle n |\hat{B}|0 \rangle,
    \end{align}
where $n$ is the total number of roots, $k$ denotes the eigenvector, and $\nu$ is the index of Jocobian matrix element. 
The above transition matrix could be written as 
    \begin{align} \label{eqn:TM}
        &\langle 0|\hat{A}|k\rangle \langle k|\hat{B}|0\rangle = \bigg\{ \langle \Lambda |[\hat{A}, \, \hat{\tau}_k]|\textrm{pCCD}\rangle - \nonumber \\ 
        & \sum_{\nu}(-\hat{J}+\omega \boldsymbol{I})_{\mu\nu}^{-1} \langle\bar{\nu}|\hat{A}|\textrm{pCCD}\rangle \langle \Lambda|[[\hat{H}_0,\hat{\tau}_\nu], \hat{\tau}_k]|\textrm{pCCD}\rangle \bigg\} \langle k |\hat{B}|0 \rangle,
    \end{align}
where $\hat{A}$ and $\hat{B}$ are arbitrary operators, $\hat{H}_0$ the Hamiltonian of Eq.~\eqref{eqn:ham}, $\hat{J}$ is the coupled cluster Jacobian, and $|\textrm{pCCD}\rangle$ denotes the pCCD wave function (in the canonical or pCCD orbital basis). 
The operators $\hat{A}$ and $\hat{B}$ can define various properties such as transition dipole moment (TDM), polarizability~\cite{pedersen-cc2-polarizability-jcp-2004, tucholska-cc-polarization-propagator-jcp-2014}, and excited density matrix (EDM), among others.
\subsection{Transition dipole moment and related properties }
The TDM, in the absence of any external potential, can be obtained using only the $\hat{A}$-dependent part of Eq.~(\ref{eqn:TM})  according to the following equation
    \begin{align} \label{eqn:TM1}
            \Gamma_{0\rightarrow k}^{\hat{A}} &= \langle 0 | \hat{A}|k\rangle \nonumber\\
            &\langle \Lambda |[\hat{A}, \, \hat{\tau}_k]|\textrm{pCCD}\rangle - \nonumber \\ 
            & \sum_{\nu}(-\hat{J}+\omega \boldsymbol{I})_{\mu\nu}^{-1} \langle\bar{\nu}|\hat{A}|\textrm{pCCD}\rangle \langle \Lambda|[[\hat{H}_0,\hat{\tau}_\nu], \hat{\tau}_k]|\textrm{pCCD}\rangle.
    \end{align}
Using dipole operators, we can obtain the TDM of selected excited states. 
An intrinsic feature of choosing pCCD as the reference wave function---in contrast to CCSD-based approaches---is that the Jacobian becomes approximately symmetric, and we can approximate transition dipole moment using only the right eigenvectors without significantly deteriorating the accuracy.
This approximation allows us to keep the computational cost to $\mathcal{O}(o^2v^2)$ (see below).
Similarly, the oscillator strength (OS) and dipole strength (DS) can be determined using Eq.~(\ref{eqn:TM1}) and the following relations,
    \begin{align} \label{eqn:os}
            \textrm{OS} &=\frac{2}{3} \omega_k \Gamma_{0\rightarrow k}^\mu \Gamma_{0\rightarrow k}^\mu,
    \end{align}
    \begin{align} \label{eqn:ds}
            \textrm{DS} &=\Gamma_{0\rightarrow k}^\mu \Gamma_{0\rightarrow k}^\mu,
    \end{align}
where \(\omega_k\) are the eigenvalues of the Jacobian (excitation energies), and \(\Gamma_{0 \rightarrow k}^\mu\) refers to the TDM elements.

\subsection{Computational scaling}
{Our pCCD-based linear response models feature a formal computational scaling of $\mathcal{O}(o^2v^2)$ (with some prefactor), neglecting the 4-index transformation of the electron repulsion integrals (ERI), which can be further reduced by working with Cholesky-decomposed ERI.
Thus, our LR-pCCD-based models are computationally more efficient than the standard LR-CCSD or EOM-CCSD models, which are of the order of $\mathcal{O}(o^2v^4)$.
Furthermore, we can store the whole Jacobian (in principle, we can also exactly diagonalize it and have access to all eigenvalues and---left and right---eigenvectors) as it scales as $\mathcal{O}(2*o^2v^2)$ in terms of memory.}

\section{Computational details}\label{sec:comput-details}
All pCCD-based calculations were performed in the developer version of the \textsc{PyBEST} v2.1.0.dev0 software package.~\cite{pybest-paper, pybest-paper-update1-cpc-2024}
The structural parameters for the small molecules (\ce{H2O}, BH, \ce{H2CO}, and furan) were taken from Refs.~\citenum{chrayteh2020mountaineering, loos2020mountaineering}.
We computed vertical excitation energies for these systems and explored different basis set variants ranging from cc-pVDZ to cc-pVQZ ~\cite{basis_dunning} {and aug-cc-pVTZ.~\mbox{\cite{aug-cc-pvtz}}}
To compare vertical excitation energies with a double electron transfer character to the CC2 and CC3 literature data, we employed the def2-TZVP basis sets.~\cite{turbomole-def2-tzvp} 
The 1 A$g^{-}$ (S0) ground state structures of all-trans polyenes were taken from Ref.~\citenum{dmrg-geom-opt}. 
For these systems, we used a cc-pVDZ basis set. 

In all our LR-pCCD+S calculations, we used the frozen core approximation (1s orbitals of C, O, and N were kept frozen) and tested two orbital sets: the canonical HF orbitals and the natural pCCD-optimized orbitals.
For the latter, we employed a variation orbital optimization protocol~\cite{oo-ap1rog, ap1rog-non-variational-orbital-optimizarion-jctc} as implemented in \textsc{PyBEST}.

The reference EOM-CCSD calculations were carried out in the \textsc{Molpro2020} software package~\cite{molpro2020-authors, molpro2020_jcp, molpro-wires} {and the reference LR-CCSD and CCSDR(3) calculations\mbox{~\cite{ccsdr3-jcp-1996}} in the \textsc{Dalton2020} software package\mbox{~\cite{dalton2013, dalton2020}}} utilizing the same basis sets and structures as in \textsc{PyBEST}.
\section{Results and Discussion}
In this section, we assess the quality of LR-pCCD+S TDMs (Eq.~\eqref{eqn:TM1}), OSs (Eq.~\eqref{eqn:os}), and DSs (SI) (Eq.~\eqref{eqn:ds}) by comparing them to other well-established quantum chemistry methods {(EOM-CCSD and LR-CCSD)} and when possible to experiment.
Our analysis is divided into three parts: (a) molecules with a dominant single electron transfer, (b) molecules with a significant admixture of double electron excitations, and (c) polymer chains.

\subsection{Single electron excitation energies}\label{sec:results-single-ee}
\begin{figure*}[ht]
    \centering
    \includegraphics[width=\textwidth]{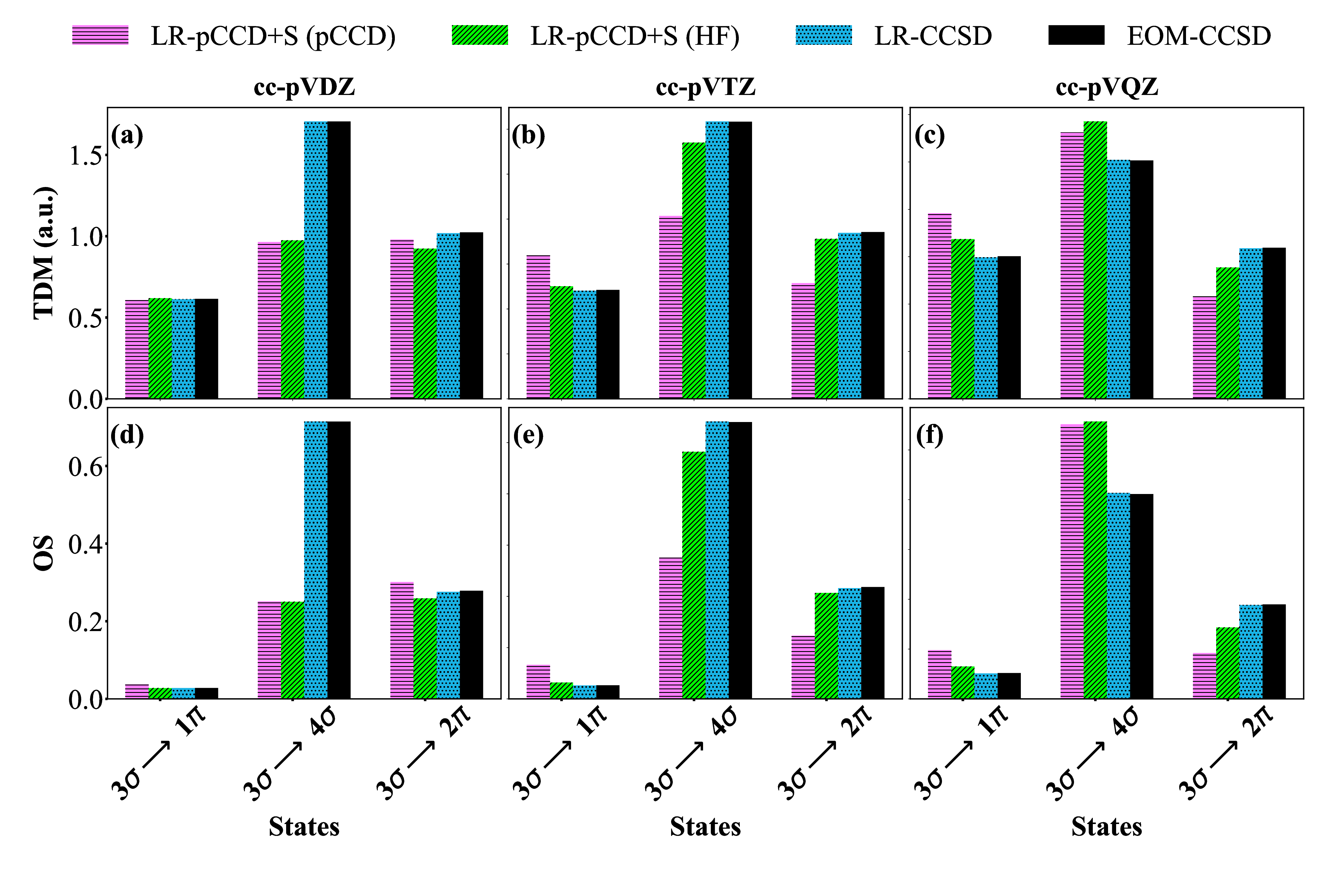}
    \caption{LR-pCCD\(+\)S(HF), LR-pCCD\(+\)S(pCCD), {LR-CCSD}, and EOM-CCSD TDM and OS for the low-lying excited states of the BH molecule using the cc-pVDZ, cc-pVTZ, and cc-pVQZ basis sets.
    The LR-pCCD\(+\)S(HF) and LR-pCCD\(+\)S(pCCD) correspond to LR-pCCD+S calculations using canonical HF and variational orbital-optimized pCCD orbitals, respectively.}
    \label{fig:bh}
\end{figure*}

\begin{figure*}[ht]
    \centering
    \includegraphics[width=\textwidth]{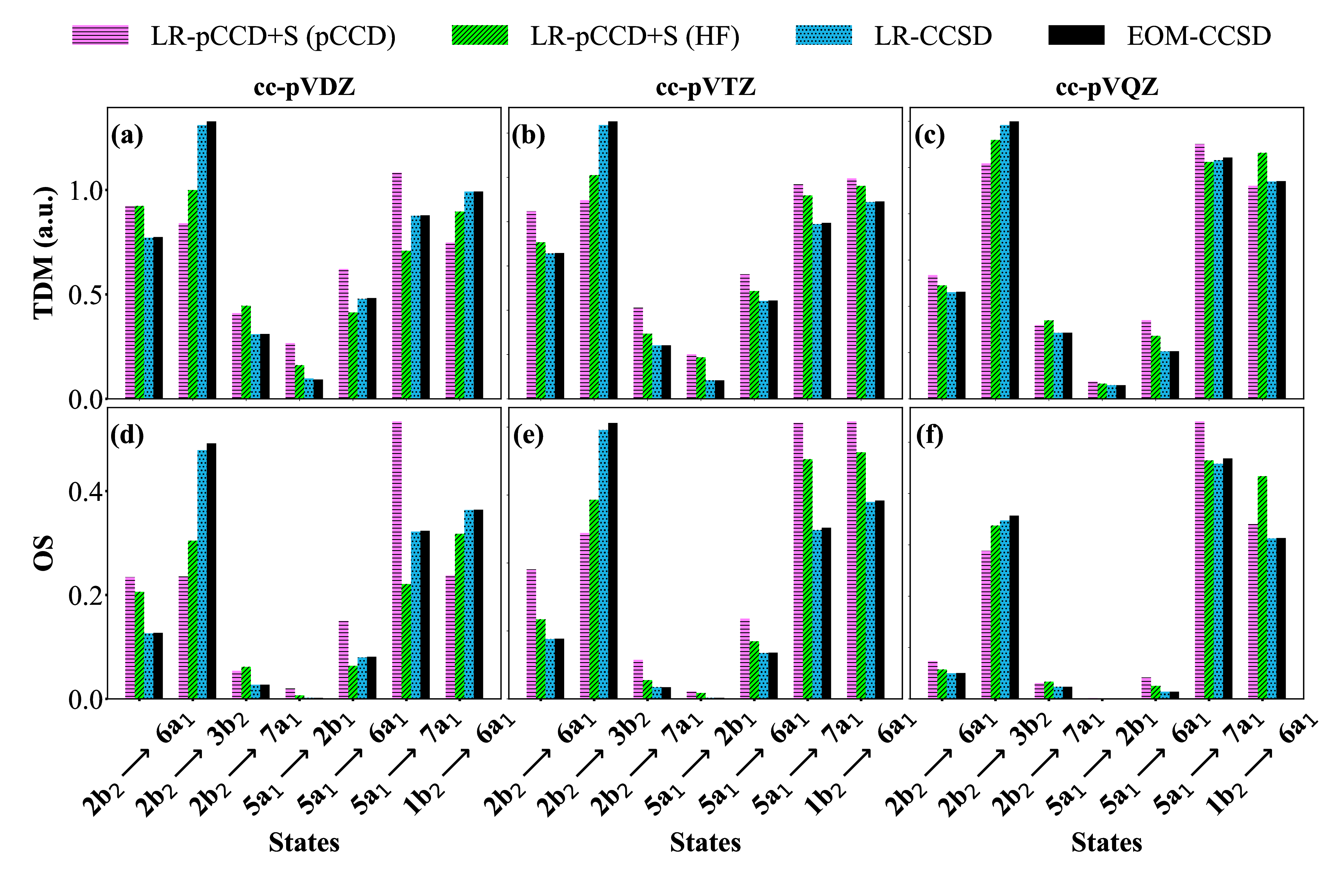}
    \caption{LR-pCCD\(+\)S(HF), LR-pCCD\(+\)S(pCCD), {LR-CCSD}, and EOM-CCSD TDM and OS for the low-lying excited states of the \ce{H2CO} molecule using the cc-pVDZ, cc-pVTZ, and cc-pVQZ basis sets.
    The LR-pCCD\(+\)S(HF) and LR-pCCD\(+\)S(pCCD) correspond to LR-pCCD+S calculations using canonical HF and variational orbital-optimized pCCD orbitals, respectively.}
    \label{fig:h2co}
\end{figure*}

\begin{figure*}[ht]
    \centering
    \includegraphics[width=\textwidth]{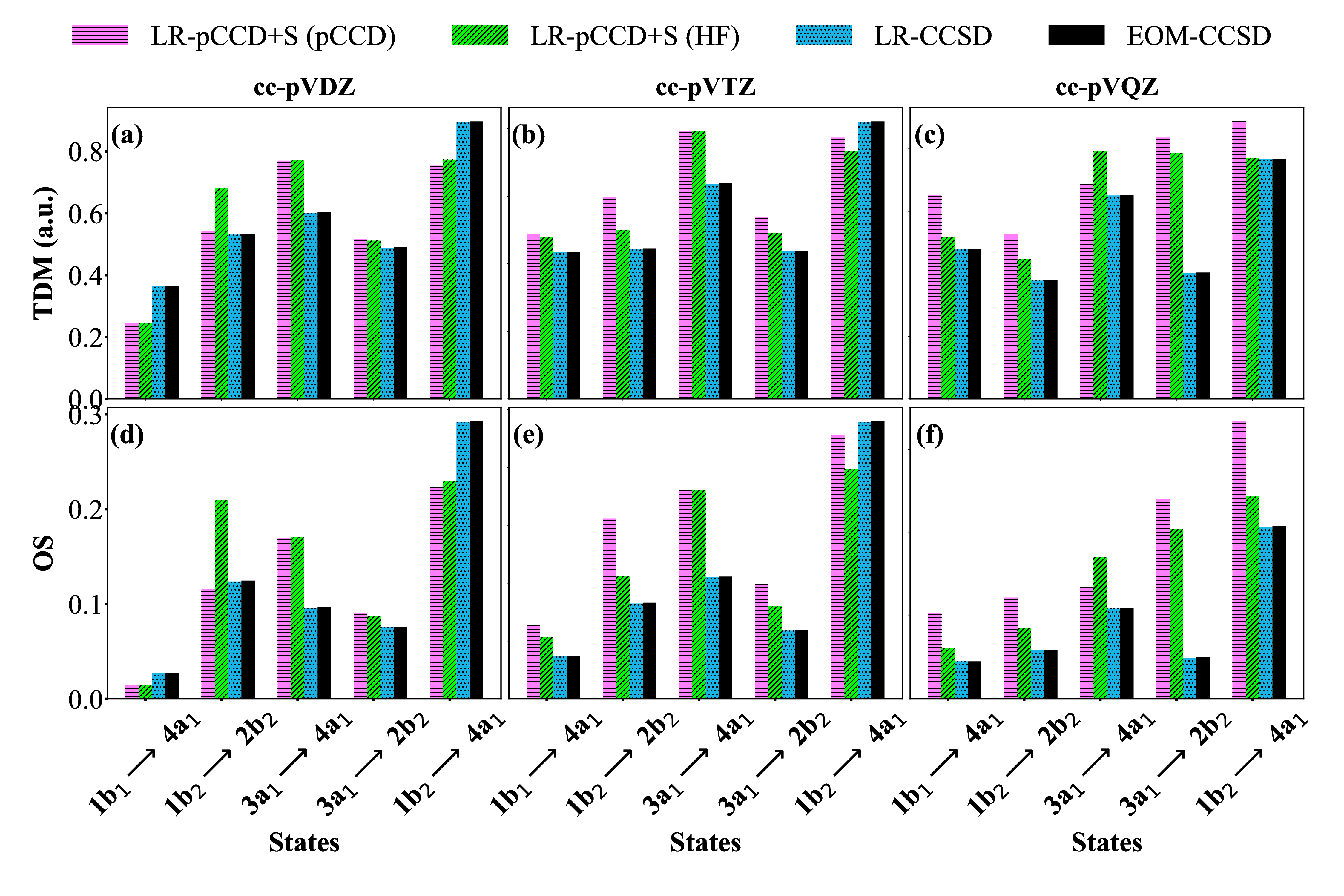}
    \caption{
    LR-pCCD\(+\)S(HF), LR-pCCD\(+\)S(pCCD), {LR-CCSD}, and EOM-CCSD TDM and OS for the low-lying excited states of the \ce{H2O} molecule using the cc-pVDZ, cc-pVTZ, and cc-pVQZ basis sets.
    The LR-pCCD\(+\)S(HF) and LR-pCCD\(+\)S(pCCD) correspond to LR-pCCD+S calculations using canonical HF and variational orbital-optimized pCCD orbitals, respectively.
    }
    \label{fig:h2o}
\end{figure*}

\begin{figure*}[ht]
    \centering
    \includegraphics[width=\textwidth]{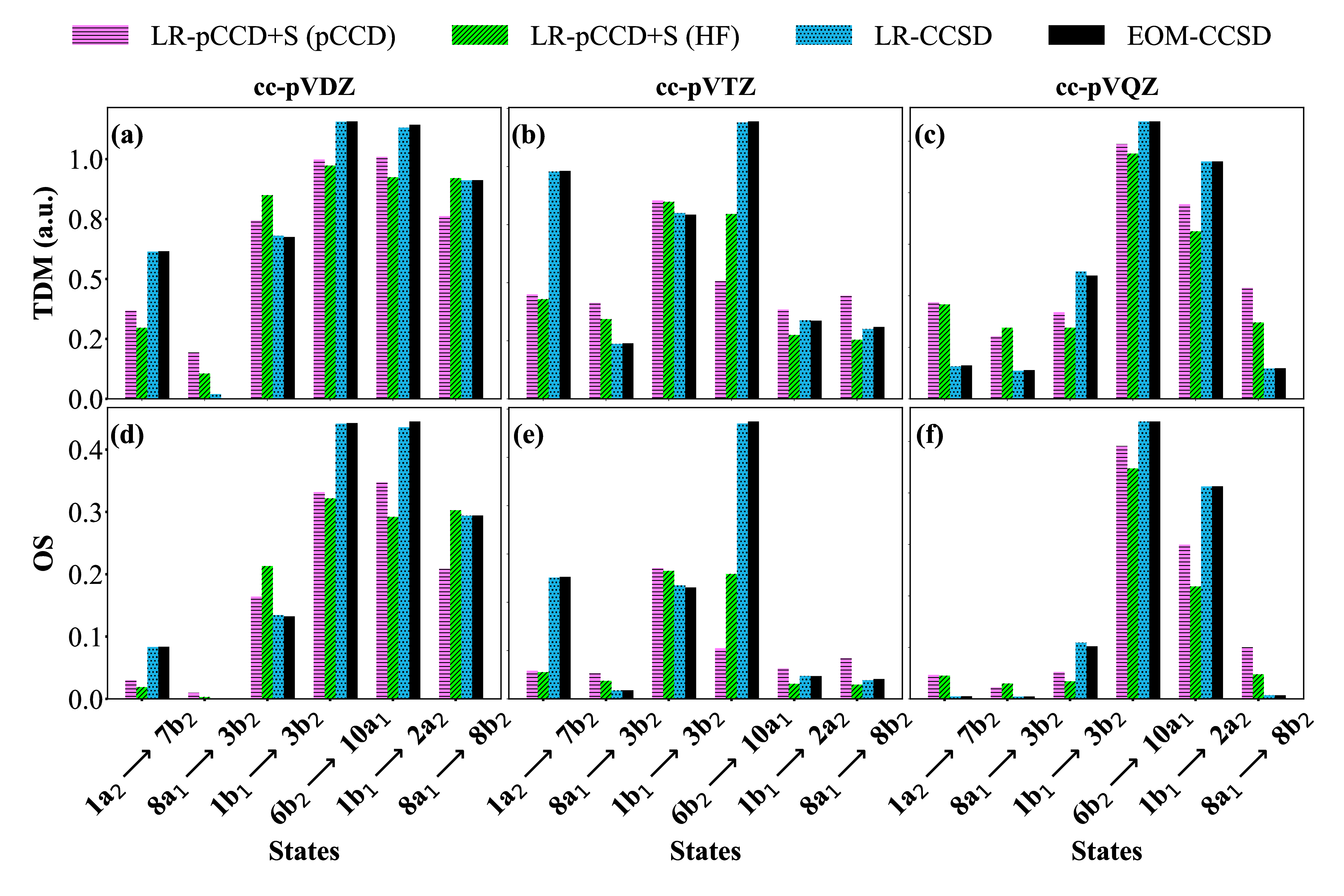}
    \caption{
    LR-pCCD\(+\)S(HF), LR-pCCD\(+\)S(pCCD), {LR-CCSD}, and EOM-CCSD TDM and OS for the low-lying excited states of the furan (\ce{C4H4O}) molecule using the cc-pVDZ, cc-pVTZ, and cc-pVQZ basis sets.
    The LR-pCCD\(+\)S(HF) and LR-pCCD\(+\)S(pCCD) correspond to LR-pCCD+S calculations using canonical HF and variational orbital-optimized pCCD orbitals, respectively.
    }
    \label{fig:furan}
\end{figure*}

We start our excited state property analysis by modeling electronic spectra using {LR-CCSD and} EOM-CCSD as reliable references, {for which, triple excitations have negligible effects (see Tables S1-S4 of the SI)}.
Specifically, we focus on the TDMs and OSs of the low-lying excited states of BH, water (\ce{H2O}), formaldehyde (\ce{H2CO}), and furan (\ce{C4H4O}) using different basis sets.
We investigate two sets of LR-pCCD+S calculations, one where we utilize the HF canonical orbitals and the other where we employ the natural pCCD orbitals (denoted as LR-pCCD+S(HF) and LR-pCCD+S(pCCD), respectively). 
The performance of LR-pCCD results for these systems is illustrated in Figures~\ref{fig:bh}--\ref{fig:furan}.
The corresponding numerical data, including excitation energies, {TDMs, OSs, and DSs,} are provided in Tables S1--S4 of the SI.

For the BH molecule, shown in Figure~\ref{fig:bh}, the first (3$\sigma \rightarrow 1\pi$) and third (3$\sigma \rightarrow 2\pi$) electronic transitions exhibit {almost the same} excited state properties, while the second transition (3$\sigma \rightarrow 4\sigma$) shows the largest values for TDM and OS.
However, when the basis set is changed from cc-pVDZ to cc-pVTZ to cc-pVQZ, {the difference in second excited state} decreases, and LR-pCCD+S {approaches LR-CCSD and} EOM-CCSD.

Similarly, a good agreement is obtained for the formaldehyde molecule, where the {canonical HF} LR-pCCD+S excited state properties agree very well with the reference {LR-CCSD and} EOM-CCSD values and trends for all investigated basis sets (cf.~Figure~\ref{fig:h2co}). 
Figure~\ref{fig:h2co} highlights the small deviations occurring between methods even for larger basis set sizes.  

We observe a somehow stronger excited-states property dependence on the basis set for the water and furan molecules shown in Figures~\ref{fig:h2o} and~\ref{fig:furan}, respectively.
Specifically, the TDMs and OSs quantitatively change when moving from the cc-pVDZ to the cc-pVTZ basis set.
Nonetheless, LR-pCCD+S predicts a qualitatively correct picture of TDMs and derived properties (OS) for all excited states.
{We should stress here that this is not always the case for the augmented basis functions (see Figure S1 of the SI)}.

A statistical analysis of the LR-pCCD+S excited state properties is summarized in Table~\ref{tab:avg_e}. 
SD and MAE are the smallest for the LR-pCCD+S(HF) approach, indicating that, on average, the canonical HF orbitals provide a slightly better basis than the natural pCCD orbitals for LR-pCCD+S excited state properties (TDMs and OSs). 
This observation aligns with what has already been observed for pCCD-based electronic excitation energies, suggesting that the ground-state pCCD orbitals might not be optimal for excited-state calculations~\cite{eom-pccd} (as to be expected).
Notably, the OS, which depends directly on the excitation energy (cf. Eq.~\ref{eqn:os}), shows no significant deviations from the statistical errors of TDMs. 
Table S5 of the SI dissects the statistical analysis by individual molecules and basis sets, providing a deeper look into their structure and basis set dependence.
The largest statistical difference (SD and MAE) between the HF and pCCD bases is observed for the {BH} molecule.

In summary, {by averaging over the basis sets and molecules summarized in Table~S5}, our in-depth comparison of the LR-pCCD+S approaches to {LR-CCSD } indicates LR-pCCD+S(HF) as the most reliable method for modeling singly excited state properties of simple molecules {(see Table\mbox{~\ref{tab:avg_e} for statistics)}}.
Changing the canonical HF basis to the natural, symmetry-broken pCCD-optimized one increases errors (SD and MAE) from about {1 to 2\%}, while the MAE for the TDM is {3 to 5\%}.
\begin{table}[ht]
\caption{Summary of statistical analysis of the LR-pCCD+S excited state properties w.r.t the {LR-CCSD} reference. SD(HF) and SD(pCCD) denote the standard deviation (SD = $\sqrt{\frac{\sum_i^N(x_i^{\rm method} - \overline{x_i})^2}{N}}$) using the canonical HF and the natural pCCD orbitals, respectively, while MAE indicates the mean absolute errors (MAE = $\sum_i^N \frac{|x_i^{\rm method} - x_i^{\rm ref}|}{N} $) using HF and pCCD orbitals. The data corresponds to all molecules, excited states, and basis sets. See also Table S5 of the SI for statistical analysis for each individual basis set. }\label{tab:avg_e}
\begin{tabular}{l|cccc}
 Property   & SD (pCCD) & SD (HF) & MAE (pCCD) & MAE (HF) \\ \hline
TDM & 0.0249 &	0.0156 & 0.0459 & 0.0327 \\
OS  & 0.0111 &	0.0083 & 0.0210 & 0.0151  \\ \hline
\end{tabular}

\end{table}

\subsection{Double electron excitations}\label{sec:results-double-ee}
In the following, we focus on electronic excitations with a non-negligible double electron transfer character and their OSs. 
Specifically, we investigate the performance of the LR-pCCD+S methods in modeling the 2 $^1$A$_1$ excited state of furan, cyclopentadiene, and pyridine. 
These are well-known, challenging examples for excited-state electronic structure methods. 
The CC2, CASSCF, and CC3 methods are often used to probe the percentage of double electron transfer in these systems.~\cite{schreiber2008benchmarks}  
As a reference for our LR-pCCD+S approaches, we use the CC3 method,~\cite{sauer2015performance} but we also report the CC2 results for comparison with our approach. 
We should stress, however, that the pure double-double excitation block of the CC2 Jacobian is only of zeroth order, meaning it is diagonal and consists solely of orbital energy differences. 
As a result, excitations dominated by singles are accurate up to the second order in CC2, whereas pure double excitations are accurate only to the zeroth order.~\cite{schreiber2008benchmarks}
Also, similar to EOM-pCCD+S, LR-pCCD+S treats the double electronic transitions approximately, which are limited to the seniority zero sector.~\cite{tamar-pccd}
Despite their questionable performance for excited states with double excitation character, we will employ CC2 and CC3 as an indication measure for the presented LR-pCCD+S methods.

Our LR-pCCD+S results are summarized in Table~\ref{tab:double_ex}.
Despite overestimating the excitation energies, LR-pCCD+S predicts oscillator strengths that agree well with the reference CC3 values for cyclopentadiene and pyridine. 
The only exception is the doubly excited state of the furan molecule.
Most likely, this doubly excited state has non-negligible contributions from broken-pair excitations, as the LR-pCCD+S-based percentage contribution of $R_2$ (more precisely, $R_\textrm{p}$) is very low compared to other methods. 
Finally, similar to what we have observed earlier, the canonical HF orbital basis provides a more reliable description of LR-pCCD+S excited states and their properties than its pCCD counterpart. 
\begin{table}[ht]
\caption{The 2\(^1\)A\(_1\) (\(\pi \rightarrow \pi^*\)) vertical excitation energies (EEs) and oscillator strengths (OSs) of furan, cyclopentadienyl, and pyridine using a def2-TZVP basis set. 
The values in parentheses represent the double excitation contributions ($R_2$ or $R_\textrm{p}$). }
\begin{tabular}{l|l|c|cc}
 & \multicolumn{1}{l|}{Molecule} &  \multicolumn{1}{c}{LR-pCCD+S(HF)} &\multicolumn{1}{c}{CC2\cite{schreiber2008benchmarks}} & \multicolumn{1}{c}{CC3\cite{sauer2015performance}($R_2$\cite{sauer2009benchmarks})}   \\ \hline
\multirow{3}{*}{EE} & \multicolumn{1}{l|}{Furan}         & 7.95 (4\%)   &  6.75  &  6.57 (16\%)      \\
                    & \multicolumn{1}{l|}{Cyclopentadiene} & 7.47 (56\%)   &  7.05  &  6.28 (22\%)     \\
                    & \multicolumn{1}{l|}{Pyridine}        & 6.44 (8\%)   &  6.88  &  6.59 (7\%)     \\ \hline
\multirow{3}{*}{OS} & \multicolumn{1}{l|}{Furan}          & 0.021  &  0.003 &  0.001   \\
                    & \multicolumn{1}{l|}{Cyclopentadiene} & 0.004 &  0.011 &  0.005  \\
                    & \multicolumn{1}{l|}{Pyridine}         & 0.024  &  0.021 &  0.014  \\ \hline
\end{tabular}\label{tab:double_ex}
\footnotetext[1]{Rate of the double excitations. }
\end{table}

\subsection{All-trans-polyenes}\label{sec:results-polyenes}
Finally, we move toward larger molecular structures to further validate our methods and investigate all-trans polyenes, C$_{2n}$H$_{2n+2}$, of different chain lengths with $n = 5, 6, 7, 8$.
Specifically, we focus on the 1 B$_u^+$ state, which is known to have a non-negligible intensity and thus to be optically active.
We used our LR-pCCD+S(HF) approach to determine the excitation energies and the corresponding OSs and TDMs.
As previously reported,~\cite{eom-pccd, eom-pccd-erratum} EOM-pCCD+S(HF) provides reliable excitation energies for the lowest-lying B$_u^+$ state in these challenging systems.
In this work, we turn our attention toward the corresponding excited state properties obtained from the proposed LR-pCCD-based models.
To the best of our knowledge, the excited state properties of longer all-trans polyenes have not yet been studied with more elaborate ab initio methods.

Our results for the S0 structures are collected in Table~\ref{tab:poly}.
For longer chains, the oscillator strength follows a decreasing trend of excitation energies of the 1B$_u^+$ state.
The TDMs and OSs reach their minima for the \ce{C14H16} molecule.
Since there is no other reliable theoretical data to verify our excited state properties, we compare our calculated excitation energies to the experimental spectra from Ref.~\citenum{christensen-ee-polyenes-jpca-2008}.
Similar to the trends in the oscillator strength predicted with LR-pCCD+S(HF), the experimental intensity decreases toward larger molecules (see Figure 2 of Ref.~\citenum{christensen-ee-polyenes-jpca-2008}).
Knowing that the LR-pCCD+S(HF) excited state properties are size intensive and the excitation energies of the 1B$_u^+$ state are reasonable (within the given basis set and molecular structures), we are quite confident that the results in Table~\ref{tab:poly} are qualitatively correct.

\begin{table}[ht]
\caption{LR-pCCD+S vertical excitation energies (EEs), oscillator strengths (OSs) and transition dipole moments (TDMs) for the B$_u^+$ state in trans-polyenes.}
\label{tab:poly}
\begin{tabular}{lccc|c}
      & \multicolumn{3}{c|}{LR-pCCD+S(HF)} & Exp.~\cite{christensen-ee-polyenes-jpca-2008} \\ \hline
    \multicolumn{1}{l|}{Molecule}  & EE (eV)   & OS & TDM   &   EE (eV) \\ \hline
    \multicolumn{1}{l|}{\ce{C10H12}} & 4.91 & 2.56  & 0.88  &  4.60  \\
    \multicolumn{1}{l|}{\ce{C12H14}} & 4.50 & 1.01   & 0.58  &  4.02 \\
    \multicolumn{1}{l|}{\ce{C14H16}} & 4.25 & 1.00 & 0.59 &  3.74   \\
    \multicolumn{1}{l|}{\ce{C16H18}} & 4.00  & 1.00 & 0.61  &  3.51  \\ \hline
\end{tabular}
\end{table}

\section{Conclusions and outlook}
In this work, we derived the working equations for the LR-pCCD and LR-pCCD+S methods.
While the first one has little use in quantum chemistry applications as it is restricted to pair excited states only, the latter provides a computationally inexpensive extension to the EOM-pCCD+S method,~\cite{eom-pccd, eom-pccd-erratum}  allowing us to target excited state properties such as TDMs, OSs, and DSs (see SI) within the pCCD model, requiring only mean-field-like computational cost. 
Most importantly, our numerical examples validate the approximate inclusion of single excitations and the restriction to the right eigenstates only.
Specifically, our study indicates a satisfyingly good performance of the LR-pCCD+S model for singly excited states, comparable to {the LR-CCSD approach and} the more expensive EOM-CCSD approach (mean-field-like vs.~$N^6$ scaling).
Similar to EOM-pCCD+S excitation energies, excited state properties within the LR-pCCD+S framework are, on average, more reliable using canonical HF orbitals than the natural pCCD-optimized orbitals. 
Moreover, we demonstrate that our LR-pCCD+S(HF) method can also be used to predict excited state properties of more complex systems with a significant amount of double/bi-excitation character.
Our LR-pCCD+S predictions are, however, limited to doubly excited states of seniority zero character. 
Finally, we provide a first theoretical description of electronic TDMs and OSs for the 1 B$_u^+$ state of longer all-trans polyene chains.
Specifically, LR-pCCD+S(HF) predicts electronic properties in good agreement with experimentally recorded spectra. 

In summary, we introduced a new cost-effective computational protocol representing a robust model to predict excited state properties qualitatively.
To reduce the errors with respect to more elaborate models and hence to reach a more quantitative level for the prediction of excited state properties, we must account for the missing dynamic correlation energy in our LR-pCCD+S approach.
Promising alternatives are various tailored coupled cluster corrections on top of the pCCD reference function~\cite{tamar-pccd,ap1rog-lcc, pccd-tcc} within the linear response framework. 
Finally, the pCCD-based transition density matrices open new ways to construct excited states' descriptors and topological analyses of charge transfer excited states~\cite{etienne2015transition} that are of great importance in organic electronics.~\cite{qc-review-dye-solar-cells-acr-2012, pccd-perspective-jpcl-2023}
\section*{Conflicts of interest}
There are no conflicts to declare.

\section{acknowledgement}
S.~A. and P.~T.~acknowledge financial support from the SONATA BIS research grant from the National Science Centre, Poland (Grant No. 2021/42/E/ST4/00302). 
Funded/Co-funded by the European Union (ERC, DRESSED-pCCD, 101077420).
Views and opinions expressed are, however, those of the author(s) only and do not necessarily reflect those of the European Union or the European Research Council. Neither the European Union nor the granting authority can be held responsible for them. 
S.~A. gratefully acknowledges Prof. Iulia Emilia Brumboiu for the very helpful discussions.

\section{suppinfo}
Tables of data for TDM and OS, and DS results from Molpro and LR-pCCD+S in HF and pCCD base calculations for BH, \ce{H2O}, \ce{H2CO}, and furan molecules across different basis sets. 
Tables of statistical calculations for SD and MAE for individual molecules in various basis sets. 
A detailed explanation of the calculations used to obtain the final relation for transition matrix elements, including the vectors and matrices referenced in Table~\ref{tab:table1}.

\section*{Data Availability Statements}
The data underlying this study are available in the published article and its Supporting Information.
The released version of the PyBEST code is available on Zenodo at \url{https://zenodo.org/records/10069179} and on PyPI at \url{https://pypi.org/project/pybest/}.

\bibliography{main_a}

\end{document}


\renewcommand{\thefigure}{S\arabic{figure}}
\renewcommand{\theequation}{S\arabic{equation}}
\renewcommand{\thesection}{S\arabic{section}}
\renewcommand{\thetable}{S\arabic{table}}
\renewcommand{\tablename}{{Table}}

\newcommand{\Tp}{{\hat{T}_{\textrm{pCCD}}}}
\thispagestyle{empty}

\begin{center}
\begin{spacing}{2.0}
{\LARGE\bf Linear response pCCD-based methods: LR-pCCD and LR-pCCD+S approaches for the efficient and reliable modelling of excited state properties}
\end{spacing}

\vspace{2cm}
{\large 
{Somayeh Ahmadkhani$^{\dagger}$, Katharina Boguslawski, and Paweł Tecmer$^{\dagger}$}
}\\[4ex]
Institute of Physics, Faculty of Physics, Astronomy, and Informatics, Nicolaus Copernicus University in Toruń, Grudzi{a}dzka 5, Toruń, 87-100 Toruń, Poland 
\\$^{\dagger}$E-mail:so.ahmadkhani@gmail.com; ptecmer@fizyka.umk.pl \\

\vspace{5cm}

{\bf \Large Supplementary Information}

\vfil

\end{center}
\newpage
\section{Numerical results for single excitation energies.}

\begin{table*}[h!]
\caption{Vertical excitation energies (EE), TDMs, DSs, and OSs for \ce{H2CO} in ground state geometry using cc-pVDZ, cc-pVTZ, cc-pVQZ and aug-cc-pVTZ basis sets.}\label{tab:si_2}
\begin{small}
\begin{tabular}{llccccccc}
 &
  \multicolumn{8}{c}{cc-pVDZ} \\ \cline{2-9} 
Property &
  Method &
  2b$_2 \longrightarrow$ 6a$_1$ &
  2b$_2 \longrightarrow$ 3b$_2$ &
  2b$_2 \longrightarrow$ 7a$_1$ &
  5a$_1 \longrightarrow$ 2b$_1$ &
  5a$_1 \longrightarrow$ 6a$_1$ &
  5a$_1 \longrightarrow$ 7a$_1$ &
  1b$_2 \longrightarrow$ 6a$_1$ \\ \hline
EE &
  LR-pCCDS(pCCD) &
  0.4121 &
  0.4999 &
  0.4817 &
  0.4160 &
  0.5826 &
  0.6829 &
  0.6371 \\
 &
  LR-pCCDS(HF) &
  0.3626 &
  0.4570 &
  0.4655 &
  0.3934 &
  0.5594 &
  0.6599 &
  0.5932 \\
 &
  LR-CCSD &
  0.3177 &
  0.4183 &
  0.4260 &
  0.3490 &
  0.5231 &
  0.6288 &
  0.5540 \\
 &
  EOM-CCSD &
  0.3177 &
  0.4183 &
  0.4260 &
  0.3490 &
  0.5231 &
  0.6288 &
  0.5540 \\
 &
  CCSDR(3) &
  0.3149 &
  0.4132 &
  0.4263 &
  0.3467 &
  0.5169 &
  0.6264 &
  0.5467 \\ \hline 
TDM &
  LR-pCCDS(pCCD) &
  0.9231 &
  0.8408 &
  0.4105 &
  0.2665 &
  0.6217 &
  1.0831 &
  0.7478 \\
 &
  LR-pCCDS(HF) &
  0.9234 &
  0.9999 &
  0.4455 &
  0.1609 &
  0.4136 &
  0.7089 &
  0.8966 \\
 &
  LR-CCSD &
  0.7716 &
  1.3105 &
  0.3095 &
  0.0975 &
  0.4783 &
  0.8766 &
  0.9925 \\
 &
  EOM-CCSD &
  0.7747 &
  1.3283 &
  0.3106 &
  0.0927 &
  0.4825 &
  0.8787 &
  0.9929 \\ \hline 
DS &
  LR-pCCDS(pCCD) &
  0.8521 &
  0.7069 &
  0.1685 &
  0.0710 &
  0.3865 &
  1.1732 &
  0.5592 \\
 &
  LR-pCCDS(HF) &
  0.8526 &
  0.9998 &
  0.1985 &
  0.0259 &
  0.1711 &
  0.5026 &
  0.8039 \\
 &
  LR-CCSD &
  0.5954 &
  1.7173 &
  0.0958 &
  0.0095 &
  0.2288 &
  0.7684 &
  0.9850 \\
 &
  EOM-CCSD &
  0.6001 &
  1.7643 &
  0.0965 &
  0.0086 &
  0.2328 &
  0.7722 &
  0.9858 \\ \hline 
OS &
  LR-pCCDS(pCCD) &
  0.2341 &
  0.2356 &
  0.0541 &
  0.0197 &
  0.1501 &
  0.5341 &
  0.2375 \\
 &
  LR-pCCDS(HF) &
  0.2061 &
  0.3046 &
  0.0616 &
  0.0068 &
  0.0638 &
  0.2211 &
  0.3179 \\
 &
  LR-CCSD &
  0.1261 &
  0.4789 &
  0.0272 &
  0.0022 &
  0.0798 &
  0.3221 &
  0.3638 \\
 &
  EOM-CCSD &
  0.1271 &
  0.4920 &
  0.0274 &
  0.0020 &
  0.0812 &
  0.3237 &
  0.3641 \\ \hline
 &
   &
  \multicolumn{7}{c}{cc-pVTZ} \\ \cline{3-9} 
 &
   &
  2b$_2 \longrightarrow$ 6a$_1$ &
  2b$_2 \longrightarrow$ 3b$_2$ &
  2b$_2 \longrightarrow$ 7a$_1$ &
  5a$_1 \longrightarrow$ 2b$_1$ &
  5a$_1 \longrightarrow$ 6a$_1$ &
  5a$_1 \longrightarrow$ 7a$_1$ &
  1b$_2 \longrightarrow$ 6a$_1$ \\ \hline
EE &
  LR-pCCDS(pCCD) &
  0.3988 &
  0.4551 &
  0.5090 &
  0.4037 &
  0.5620 &
  0.6510 &
  0.6176 \\
 &
  LR-pCCDS(HF) &
  0.3515 &
  0.4309 &
  0.4783 &
  0.3715 &
  0.5363 &
  0.6268 &
  0.5888 \\
 &
  LR-CCSD &
  0.3065 &
  0.3881 &
  0.4408 &
  0.3435 &
  0.5180 &
  0.5983 &
  0.5508 \\
 &
  EOM-CCSD &
  0.3065 &
  0.3881 &
  0.4408 &
  0.3435 &
  0.5180 &
  0.5983 &
  0.5508 \\
 &
  CCSDR(3) &
  0.3025 &
  0.3829 &
  0.4366 &
  0.3406 &
  0.5146 &
  0.5933 &
  0.5450 \\ \hline 
TDM &
  LR-pCCDS(pCCD) &
  0.8467 &
  0.8972 &
  0.4116 &
  0.2012 &
  0.5617 &
  0.9677 &
  0.9958 \\
 &
  LR-pCCDS(HF) &
  0.7069 &
  1.0101 &
  0.2943 &
  0.1873 &
  0.4870 &
  0.9186 &
  0.9615 \\
 &
  LR-CCSD &
  0.6563 &
  1.2365 &
  0.2419 &
  0.0837 &
  0.4422 &
  0.7892 &
  0.8885 \\
 &
  EOM-CCSD &
  0.6585 &
  1.2528 &
  0.2419 &
  0.0837 &
  0.4441 &
  0.7947 &
  0.8916 \\ \hline 
DS &
  LR-pCCDS(pCCD) &
  0.7169 &
  0.8049 &
  0.1694 &
  0.0405 &
  0.3155 &
  0.9364 &
  0.9917 \\
 &
  LR-pCCDS(HF) &
  0.4997 &
  1.0203 &
  0.0866 &
  0.0351 &
  0.2372 &
  0.8438 &
  0.9245 \\
 &
  LR-CCSD &
  0.4307 &
  1.5290 &
  0.0585 &
  0.0070 &
  0.1955 &
  0.6228 &
  0.7895 \\
 &
  EOM-CCSD &
  0.4336 &
  1.5696 &
  0.0585 &
  0.0070 &
  0.1972 &
  0.6315 &
  0.7949 \\ \hline 
OS &
  LR-pCCDS(pCCD) &
  0.1906 &
  0.2442 &
  0.0575 &
  0.0109 &
  0.1182 &
  0.4064 &
  0.4083 \\
 &
  LR-pCCDS(HF) &
  0.1171 &
  0.2931 &
  0.0276 &
  0.0087 &
  0.0848 &
  0.3526 &
  0.3629 \\
 &
  LR-CCSD &
  0.0880 &
  0.3956 &
  0.0172 &
  0.0016 &
  0.0675 &
  0.2484 &
  0.2899 \\
 &
  EOM-CCSD &
  0.0886 &
  0.4061 &
  0.0172 &
  0.0016 &
  0.0681 &
  0.2519 &
  0.2919 \\ \hline
 &
   &
  \multicolumn{7}{c}{cc-pVQZ} \\ \cline{3-9} 
 &
   &
  2b$_2 \longrightarrow$ 6a$_1$ &
  2b$_2 \longrightarrow$ 3b$_2$ &
  2b$_2 \longrightarrow$ 7a$_1$ &
  5a$_1 \longrightarrow$ 2b$_1$ &
  5a$_1 \longrightarrow$ 6a$_1$ &
  5a$_1 \longrightarrow$ 7a$_1$ &
  1b$_2 \longrightarrow$ 6a$_1$ \\ \hline
EE &
  LR-pCCDS(pCCD) &
  0.3861 &
  0.4171 &
  0.4518 &
  0.4006 &
  0.5395 &
  0.6680 &
  0.6022 \\
 &
  LR-pCCDS(HF) &
  0.3568 &
  0.4039 &
  0.4383 &
  0.3835 &
  0.5130 &
  0.6642 &
  0.5755 \\
 &
  LR-CCSD &
  0.3529 &
  0.3722 &
  0.4326 &
  0.3426 &
  0.4992 &
  0.6455 &
  0.5307 \\
 &
  EOM-CCSD &
  0.3529 &
  0.3722 &
  0.4326 &
  0.3426 &
  0.4992 &
  0.6455 &
  0.5307 \\
 &
  CCSDR(3) &
  0.3507 &
  0.3665 &
  0.4284 &
  0.3394 &
  0.4923 &
  0.6385 &
  0.5223 \\
  \hline 
TDM &
  LR-pCCDS(pCCD) &
  0.5351 &
  1.0182 &
  0.3187 &
  0.0748 &
  0.3404 &
  1.1017 &
  0.9212 \\
 &
  LR-pCCDS(HF) &
  0.4904 &
  1.1196 &
  0.3391 &
  0.0656 &
  0.2726 &
  1.0242 &
  1.0633 \\
 &
  LR-CCSD &
  0.4610 &
  1.1839 &
  0.2855 &
  0.0592 &
  0.2052 &
  1.0316 &
  0.9398 \\
 &
  EOM-CCSD &
  0.4628 &
  1.1995 &
  0.2860 &
  0.0592 &
  0.2059 &
  1.0433 &
  0.9412 \\
  \hline 
DS &
  LR-pCCDS(pCCD) &
  0.2863 &
  1.0368 &
  0.1016 &
  0.0056 &
  0.1159 &
  1.2137 &
  0.8486 \\
 &
  LR-pCCDS(HF) &
  0.2405 &
  1.2534 &
  0.1150 &
  0.0043 &
  0.0743 &
  1.0490 &
  1.1307 \\
 &
  LR-CCSD &
  0.2125 &
  1.4017 &
  0.0815 &
  0.0035 &
  0.0421 &
  1.0643 &
  0.8833 \\
 &
  EOM-CCSD &
  0.2142 &
  1.4387 &
  0.0818 &
  0.0035 &
  0.0424 &
  1.0885 &
  0.8858 \\ \hline 
OS &
  LR-pCCDS(pCCD) &
  0.0737 &
  0.2883 &
  0.0306 &
  0.0015 &
  0.0417 &
  0.5405 &
  0.3407 \\
 &
  LR-pCCDS(HF) &
  0.0572 &
  0.3375 &
  0.0336 &
  0.0011 &
  0.0254 &
  0.4645 &
  0.4338 \\
 &
  LR-CCSD &
  0.0500 &
  0.3478 &
  0.0235 &
  0.0008 &
  0.0140 &
  0.4580 &
  0.3125 \\
 &
  EOM-CCSD &
  0.0504 &
  0.3570 &
  0.0236 &
  0.0008 &
  0.0141 &
  0.4684 &
  0.3134 \\ \hline
  \end{tabular}
  \end{small}
\end{table*}
\FloatBarrier
\clearpage
\thispagestyle{mypagestyle1}  
\begin{table}
\centering
\renewcommand{\arraystretch}{0.83}
\setlength\tabcolsep{6pt} 
\DefTblrTemplate{firsthead, middlehead,lasthead}{default}{}
 \begin{small}
\begin{tabular}{l|l|ccccccc}
   &
   
  \multicolumn{7}{c}{aug-cc-pVTZ} \\  \hline \\
 Property & Method &
  2b$_2 \longrightarrow$ 6a$_1$ &
  2b$_2 \longrightarrow$ 3b$_2$ &
  2b$_2 \longrightarrow$ 7a$_1$ &
  5a$_1 \longrightarrow$ 2b$_1$ &
  5a$_1 \longrightarrow$ 6a$_1$ &
  5a$_1 \longrightarrow$ 7a$_1$ &
  1b$_2 \longrightarrow$ 6a$_1$ \\ 
  \hline \\
EE & 												
  LR-pCCDS(pCCD) &
  0.3477 &
  0.3961 &
  0.4499 &
  0.3768 &
  0.4150 &
  0.5685 &
  0.5555 \\ 
 &												
  LR-pCCDS(HF) &
  0.3207 &
  0.3583 &
  0.4228 &
  0.3507 &
  0.4204 &
  0.5619 &
  0.5487 \\ 
 &												
  LR-CCSD &
  0.2984 &
  0.3554 &
  0.4031 &
  0.3411 &
  0.3805 &
  0.5377 &
  0.3825 \\ 
 & 												
  EOM-CCSD &
  0.2984 &
  0.3554 &
  0.4031 &
  0.3411 &
  0.3805 &
  0.5377 &
  0.3825 \\ 
 & 												
  CCSDR(3) &
  0.2970 &
  0.3495 &
  0.4001 &
  0.3381 &
  0.3788 &
  0.5283 &
  0.3811 \\ \\
  \hline \\
TDM &
  LR-pCCDS(pCCD) &
  0.4237 &
  1.0166 &
  0.6748 &
  0.1828 &
  0.2681 &
  0.8889 &
  0.7575 \\ 
 &
  LR-pCCDS(HF) &
  0.5182 &
  1.2075 &
  0.5809 &
  0.0583 &
  0.1400 &
  0.4179 &
  0.6158 \\ 
 &
  LR-CCSD &
  0.4478 &
  0.7669 &
  0.4050 &
  0.0520 &
  0.0100 &
  0.2200 &
  0.1997 \\ 
 &
  EOM-CCSD &
  0.4496 &
  0.7780 &
  0.4063 &
  0.0548 &
  0.0100 &
  0.2205 &
  0.2005 \\ \\
  \hline \\
DS &
  LR-pCCDS(pCCD) &
  0.1795 &
  1.0335 &
  0.4554 &
  0.0334 &
  0.0719 &
  0.7902 &
  0.5738 \\ 
 &
  LR-pCCDS(HF) &
  0.2685 &
  1.4581 &
  0.3374 &
  0.0034 &
  0.0196 &
  0.1746 &
  0.3792 \\ 
 &
  LR-CCSD &
  0.2005 &
  0.5881 &
  0.1640 &
  0.0027 &
  0.0001 &
  0.0484 &
  0.0399 \\ 
 &
  EOM-CCSD &
  0.2021 &
  0.6053 &
  0.1651 &
  0.0030 &
  0.0001 &
  0.0486 &
  0.0402 \\ \\
  \hline \\
OS &
  LR-pCCDS(pCCD) &
  0.0416 &
  0.2729 &
  0.1366 &
  0.0084 &
  0.0199 &
  0.2995 &
  0.2125 \\ 
 &
  LR-pCCDS(HF) &
  0.0574 &
  0.3483 &
  0.0951 &
  0.0008 &
  0.0055 &
  0.0654 &
  0.1387 \\ 
 &
  LR-CCSD &
  0.0399 &
  0.1393 &
  0.0441 &
  0.0006 &
  0.0000 &
  0.0174 &
  0.0102 \\ 
 &
  EOM-CCSD &
  0.0402 &
  0.1434 &
  0.0444 &
  0.0007 &
  0.0000 &
  0.0174 &
  0.0103 \\  \\ \hline \\
\end{tabular}
\end{small}
\end{table}
\clearpage
\begin{table*}
\caption{Vertical excitation energies (EE), TDMs, DSs, and OSs for BH in ground state geometry using cc-pVDZ, cc-pVTZ,  cc-pVQZ and aug-cc-pVTZ basis sets.}\label{tab:si_1}
\begin{small}
\begin{tabular}{llccc|ccc}
    & \multicolumn{4}{c|}{cc-pVDZ}                    & \multicolumn{3}{c}{cc-pVTZ}   \\ \cline{2-8} 
Property &
  Method &
  3$\sigma \longrightarrow$ 1$\pi$ &
  3$\sigma \longrightarrow$ 4$\sigma$ &
  3$\sigma \longrightarrow$ 2$\pi$ &
  3$\sigma \longrightarrow$ 1$\pi$ &
  3$\sigma \longrightarrow$ 4$\sigma$ &
  3$\sigma \longrightarrow$ 2$\pi$ \\ \hline
EE  & LR-pCCD+S (pCCD) & 0.1527   & 0.4052  & 0.4705  & 0.1563      & 0.3995 & 0.4467 \\
    & LR-pCCD+S (HF)   & 0.1089   & 0.3951  & 0.4550  & 0.1223      & 0.3559 & 0.3917 \\
    & LR-CCSD          & 0.1119   & 0.3683  & 0.3987  & 0.1089      & 0.3408 & 0.3803 \\
    & EOM-CCSD         & 0.1119   & 0.3683  & 0.3987  & 0.1086      & 0.3408 & 0.3803 \\
    & CCSDR(3)         & 0.1111   & 0.3673  & 0.3949  & 0.1076      & 0.3392 & 0.3772 \\ \hline
TDM & LR-pCCD+S (pCCD) & 0.6078   & 0.9645  & 0.9797  & 0.7989      & 1.0176 & 0.6437 \\
    & LR-pCCD+S (HF)   & 0.6188   & 0.9744  & 0.9239  & 0.6265      & 1.4256 & 0.8899 \\
    & LR-CCSD          & 0.6126   & 1.7061  & 1.0190  & 0.6019      & 1.5435 & 0.9233 \\
    & EOM-CCSD         & 0.6148   & 1.7058  & 1.0240  & 0.6061      & 1.5418 & 0.9279 \\ \hline
DS  & LR-pCCD+S (pCCD) & 0.3694   & 0.9303  & 0.9599  & 0.6382      & 1.0355 & 0.4144 \\
    & LR-pCCD+S (HF)   & 0.3829   & 0.9495  & 0.8535  & 0.3925      & 2.0323 & 0.7919 \\
    & LR-CCSD          & 0.3753   & 2.9108  & 1.0384  & 0.3623      & 2.3825 & 0.8524 \\
    & EOM-CCSD         & 0.3780   & 2.9096  & 1.0485  & 0.3674      & 2.3772 & 0.8610  \\ \hline
OS  & LR-pCCD+S (pCCD) & 0.0376   & 0.2513  & 0.3011  & 0.0665      & 0.2758 & 0.1234 \\
    & LR-pCCD+S (HF)   & 0.0278   & 0.2501  & 0.2589  & 0.0320      & 0.4822 & 0.2068 \\
    & LR-CCSD          & 0.0280   & 0.7147  & 0.2760  & 0.0263      & 0.5413 & 0.2161 \\
    & EOM-CCSD         & 0.0282   & 0.7144  & 0.2787  & 0.0266      & 0.5401 & 0.2183 \\ \hline
    &                   &\multicolumn{3}{c|}{cc-pVQZ} &\multicolumn{3}{c}{aug-cc-pVTZ}       \\ \hline \\
 Property & Method &
  3$\sigma \longrightarrow$ 1$\pi$ &
  3$\sigma \longrightarrow$ 4$\sigma$ &
  3$\sigma \longrightarrow$ 2$\pi$ &
  3$\sigma \longrightarrow$ 1$\pi$ &
  3$\sigma \longrightarrow$ 4$\sigma$ &
  3$\sigma \longrightarrow$ 2$\pi$ \\ \hline
EE  & LR-pCCD+S (pCCD) & 0.1570   & 0.3769  & 0.4030  &   0.1590 &
  0.2767 &
   0.3150 \\
    & LR-pCCD+S (HF)   & 0.1197   & 0.3272  & 0.3708  &  0.1161 &
   0.2447 &
   0.2839 \\ 
    & LR-CCSD          & 0.1076   & 0.3046  & 0.3498  &    0.1075 &
  0.2405  &
  0.2808  \\
    & EOM-CCSD         & 0.1076   & 0.3046  & 0.3498  &    0.1075  &
  0.2405  &
  0.2808  \\
    & CCSDR(3)         & 0.1065   & 0.3028  & 0.3476  & 0.1052  &
   0.2391 &
   0.2796  \\ \hline
TDM & LR-pCCD+S (pCCD) & 0.7836   & 1.1245  & 0.4319  & 1.1948      & 2.034  & 0.564  \\
    & LR-pCCD+S (HF)   & 0.6742   & 1.1713  & 0.5545  & 2.3298      & 2.0032 & 0.7479 \\
    & LR-CCSD          & 0.5974   & 1.0092  & 0.6359  & 2.2403      & 2.0857 & 0.7135 \\
    & EOM-CCSD         & 0.6021   & 1.0065  & 0.6379  & 2.2272      & 2.0798 & 0.7142 \\ \hline
DS  & LR-pCCD+S (pCCD) & 0.6140   & 1.2644  & 0.1865  & 1.4276      & 4.1371 & 0.3181 \\
    & LR-pCCD+S (HF)   & 0.4545   & 1.3720  & 0.3075  & 5.4279      & 4.0129 & 0.5593 \\
    & LR-CCSD          & 0.3569   & 1.0184  & 0.4044  & 5.0191      & 4.3503 & 0.5091 \\
    & EOM-CCSD         & 0.3625   & 1.0130  & 0.4069  & 4.9605      & 4.3254 & 0.5101 \\ \hline
OS  & LR-pCCD+S (pCCD) & 0.0490   & 0.2758  & 0.0461  & 0.1105      & 0.6749 & 0.0602 \\
    & LR-pCCD+S (HF)   & 0.0326   & 0.2786  & 0.0717  & 0.3890      & 0.6434 & 0.1047 \\
    & LR-CCSD          & 0.0256   & 0.2068  & 0.0943  & 0.3597      & 0.6975 & 0.0953 \\
    & EOM-CCSD         & 0.0260   & 0.2057  & 0.0949  & 0.3555      & 0.6935 & 0.0955\\ \hline
\end{tabular}  
\end{small}
\end{table*}
\begin{table*}
\begin{center}
\caption{Vertical excitation energies (EE), TDMs, DSs, and OSs for \ce{H2O} in ground state geometry using cc-pVDZ, cc-pVTZ, cc-pVQZ and aug-cc-pVTZ basis sets.}\label{tab:si_3}
\begin{tabular}{llccccc}
\multicolumn{7}{c}{cc-pVDZ} \\ \hline
Property & Method &
  1b$_1 \longrightarrow$ 4a$_1$ &
  1b$_2 \longrightarrow$ 2b$_2$ &
  3a$_1 \longrightarrow$ 4a$_1$ &
  3a$_1 \longrightarrow$ 2b$_2$ &
  1b$_2 \longrightarrow$ 4a$_1$ \\ \hline
EE &
  LR-pCCD+S (pCCD) & 0.3630 &0.5908 & 0.4301 &0.5131 & 0.5903 \\
 &
  LR-pCCD+S (HF) & 0.3539 &0.6751 &0.4286 &0.5036 &0.5779 \\
 &
  LR-CCSD & 0.3001 & 0.6584 & 0.3976 & 0.4746 & 0.5454 \\
 & EOM-CCSD & 0.3001 & 0.6584 & 0.3976 & 0.4746 & 0.5454 \\
 &
  CCSDR(3) & 0.3016 & 0.6571 & 0.3986 & 0.4753 & 0.5448 \\
TDM &
  LR-pCCD+S (pCCD) &
  0.2447 &
  0.5430 &
  0.7700 &
  0.5158 &
  0.7729 \\
 &
  LR-pCCD+S (HF) &
  0.2454 &
  0.6821 &
  0.7727 &
  0.5114 &
  0.7728 \\
 &
  LR-CCSD &
  0.3661 &
  0.5310 &
  0.6018 &
  0.4885 &
  0.8963 \\
 &
  EOM-CCSD &
  0.3661 &
  0.5330 &
  0.6034 &
  0.4898 &
  0.8969 \\
DS &
  LR-pCCD+S (pCCD) &
  0.0599 &
  0.2948 &
  0.5929 &
  0.2660 &
  0.5974 \\
 &
  LR-pCCD+S (HF) &
  0.0602 &
  0.4653 &
  0.5971 &
  0.2615 &
  0.5972 \\
 &
  LR-CCSD &
  0.1340 &
  0.2820 &
  0.3622 &
  0.2386 &
  0.8034 \\
 &
  EOM-CCSD &
  0.1340 &
  0.2841 &
  0.3641 &
  0.2399 &
  0.8045 \\
OS &
  LR-pCCD+S (PCCD) &
  0.0145 &
  0.1161 &
  0.1700 &
  0.0910 &
  0.2351 \\
 &
  LR-pCCD+S (HF) &
  0.0142 &
  0.2094 &
  0.1706 &
  0.0878 &
  0.2301 \\
 &
  LR-CCSD &
  0.0268 &
  0.1238 &
  0.0960 &
  0.0755 &
  0.2921 \\
 &
  EOM-CCSD &
  0.0268 &
  0.1247 &
  0.0965 &
  0.0759 &
  0.2925 \\ \hline
 &
  \multicolumn{6}{c}{cc-pVTZ} \\ \cline{2-7} 
  &
   &
  1b$_1 \longrightarrow$ 4a$_1$ &
  1b$_2 \longrightarrow$ 2b$_2$ &
  3a$_1 \longrightarrow$ 4a$_1$ &
  3a$_1 \longrightarrow$ 2b$_2$ &
  1b$_2 \longrightarrow$ 4a$_1$ \\ \hline
EE  & LR-pCCD+S (pCCD) & 0.3988 & 0.6510 & 0.4295 & 0.5090 & 0.5706 \\
    & LR-pCCD+S (HF)   & 0.3486 & 0.6364 & 0.4290 & 0.5025 & 0.5540 \\
    & LR-CCSD          & 0.2977 & 0.6304 & 0.3895 & 0.4652 & 0.5328 \\
    & EOM-CCSD         & 0.2977 & 0.6304 & 0.3895 & 0.4652 & 0.5328 \\
    & CCSDR(3)         & 0.2979 & 0.6292 & 0.3894 & 0.4652 & 0.5321 \\ \hline
TDM & LR-pCCD+S (pCCD) & 0.488  & 0.599  & 0.794  & 0.5391 & 0.7737 \\
    & LR-pCCD+S (HF)   & 0.478  & 0.5001 & 0.7938 & 0.4902 & 0.7333 \\
    & LR-CCSD          & 0.4341 & 0.4428 & 0.6359 & 0.4361 & 0.8204 \\
    & EOM-CCSD         & 0.4335 & 0.4446 & 0.638  & 0.4384 & 0.8215 \\ \hline
DS  & LR-pCCD+S (pCCD) & 0.2381 & 0.3588 & 0.6304 & 0.2906 & 0.5986 \\
    & LR-pCCD+S (HF)   & 0.2285 & 0.2501 & 0.6301 & 0.2403 & 0.5377 \\
    & LR-CCSD          & 0.1884 & 0.1961 & 0.4044 & 0.1902 & 0.6731 \\
    & EOM-CCSD         & 0.1879 & 0.1977 & 0.4071 & 0.1922 & 0.6748 \\ \hline
OS  & LR-pCCD+S (PCCD) & 0.0633 & 0.1557 & 0.1805 & 0.0986 & 0.2277 \\
    & LR-pCCD+S (HF)   & 0.0531 & 0.1061 & 0.1802 & 0.0805 & 0.1986 \\
    & LR-CCSD          & 0.0374 & 0.0824 & 0.1050 & 0.0590 & 0.2391 \\
    & EOM-CCSD         & 0.0373 & 0.0831 & 0.1057 & 0.0596 & 0.2397 \\ \hline
 &
  \multicolumn{6}{c}{cc-pVQZ} \\ \cline{2-7} 
 &
   &
  1b$_1 \longrightarrow$ 4a$_1$ &
  1b$_2 \longrightarrow$ 2b$_2$ &
  3a$_1 \longrightarrow$ 4a$_1$ &
  3a$_1 \longrightarrow$ 2b$_2$ &
  1b$_2 \longrightarrow$ 4a$_1$ \\ \hline
EE  & LR-pCCD+S (pCCD) & 0.3624 & 0.6563 & 0.4270 & 0.5162 & 0.6353 \\
    & LR-pCCD+S (HF)   & 0.3417 & 0.6383 & 0.4068 & 0.4935 & 0.6158 \\
    & LR-CCSD          & 0.2949 & 0.6121 & 0.3847 & 0.4582 & 0.5274 \\
    & EOM-CCSD         & 0.2949 & 0.6121 & 0.3847 & 0.4582 & 0.5274 \\
    & CCSDR(3)         & 0.2944 & 0.6111 & 0.3840 & 0.4579 & 0.5264 \\ \hline
TDM & LR-pCCD+S (pCCD) & 0.6529 & 0.5280 & 0.6861 & 0.8363 & 0.8879 \\
    & LR-pCCD+S (HF)   & 0.5187 & 0.4472 & 0.7929 & 0.7878 & 0.7714 \\
    & LR-CCSD          & 0.4795 & 0.3785 & 0.6510 & 0.4021 & 0.7677 \\
    & EOM-CCSD         & 0.4794 & 0.3796 & 0.6531 & 0.4042 & 0.7685 \\ \hline
DS  & LR-pCCD+S (pCCD) & 0.4263 & 0.2788 & 0.4707 & 0.6994 & 0.7884 \\
    & LR-pCCD+S (HF)   & 0.2691 & 0.2000 & 0.6287 & 0.6207 & 0.5951 \\
    & LR-CCSD          & 0.2299 & 0.1433 & 0.4238 & 0.1617 & 0.5893 \\
    & EOM-CCSD         & 0.2298 & 0.1441 & 0.4266 & 0.1634 & 0.5906 \\ \hline
OS  & LR-pCCD+S (pCCD) & 0.1030 & 0.1220 & 0.1340 & 0.2407 & 0.3339 \\
    & LR-pCCD+S (HF)   & 0.0613 & 0.0851 & 0.1705 & 0.2042 & 0.2443 \\
    & LR-CCSD          & 0.0452 & 0.0585 & 0.1087 & 0.0494 & 0.2072 \\
    & EOM-CCSD         & 0.0452 & 0.0588 & 0.1094 & 0.0499 & 0.2077 \\ \hline
 &
 \end{tabular}
\end{center}
\end{table*}

\FloatBarrier
\clearpage
\thispagestyle{mypagestyle2}  

\begin{table}
\centering
\renewcommand{\arraystretch}{0.83}
\setlength\tabcolsep{6pt} 
\DefTblrTemplate{firsthead, middlehead,lasthead}{default}{}
\begin{tabular}{llccccc}
  \multicolumn{6}{c}{aug-cc-pVTZ} \\ \hline \\
 Property &
  Method &
  1b$_1 \longrightarrow$ 4a$_1$ &
  1b$_2 \longrightarrow$ 2b$_2$ &
  3a$_1 \longrightarrow$ 4a$_1$ &
  3a$_1 \longrightarrow$ 2b$_2$ &
  1b$_2 \longrightarrow$ 4a$_1$ \\ \hline \\
EE &							
  LR\_pCCDS(pCCD) &
  0.3480  &
  0.4605  &
  0.4093  &
  0.4789 &
  0.5377 \\
 &								
  LR\_pCCDS(HF) &
  0.3261 &
  0.4402 &
  0.3872 &
  0.4602 &
  0.5113 \\
 &								
  LR-CCSD &
  0.2792 &
  0.4175 &
  0.3659 &
  0.4298 &
  0.4878 \\
 &							
  EOM-CCSD &
  0.2792 &
  0.4175 &
  0.3659 &
  0.4298 &
  0.4957 \\
 &								
  CCSDR(3) &
  0.2792 &
  0.4177 &
  0.3659 &
  0.4305 &
  0.4885 \\ \\ \hline \\
TDM &
  LR\_pCCDS(pCCD) &
  0.7712 &
  0.1411 &
  0.7835 &
  0.4199 &
  0.3111 \\
 &
  LR\_pCCDS(HF) &
  0.7424 &
  0.1196 &
  0.7407 &
  0.3111 &
  0.4970 \\
 &
  LR-CCSD &
  0.5362 &
  0.0200 &
  0.6346 &
  0.2415 &
  0.4223 \\
 &
  EOM-CCSD &
  0.5364 &
  0.0600 &
  0.6365 &
  0.2425 &
  0.4268 \\ \\ \hline \\
DS &
  LR\_pCCDS(pCCD) &
  0.5948 &
  0.0199 &
  0.6139 &
  0.1763 &
  0.0968 \\
 &
  LR\_pCCDS(HF) &
  0.5511 &
  0.0143 &
  0.5486 &
  0.0968 &
  0.2470 \\
 &
  LR-CCSD &
  0.2875 &
  0.0004 &
  0.4027 &
  0.0583 &
  0.1783 \\
 &
  EOM-CCSD &
  0.2877 &
  0.0036 &
  0.4051 &
  0.0588 &
  0.1822 \\ \\ \hline \\
OS &
  LR\_pCCDS(pCCD) &
  0.1380 &
  0.0061 &
  0.1675 &
  0.0563 &
  0.0347 \\
 &
  LR\_pCCDS(HF) &
  0.1198 &
  0.0042 &
  0.1416 &
  0.0297 &
  0.0842 \\
 &
  LR-CCSD &
  0.0535 &
  0.0001 &
  0.0982 &
  0.0167 &
  0.0580 \\
 &
  EOM-CCSD &
  0.0535 &
  0.0010 &
  0.0988 &
  0.0168 &
  0.0602 \\ \\ \hline \\
\end{tabular}
\end{table}
\clearpage

\begin{table*}
\caption{Vertical excitation energies (EE), TDMs, DSs, and OSs for furan in ground state geometry using cc-pVDZ, cc-pVTZ, cc-pVQZ and aug-cc-pVTZ basis sets.}\label{tab:si_4}
\begin{tabular}{llcccccc}
 &
  \multicolumn{7}{c}{cc-pVDZ} \\ \cline{2-8} 
Property &
  Method &
  1a$_2 \longrightarrow$ 7b$_2$ &
  8a$_1 \longrightarrow$ 3b$_2$ &
  1b$_1 \longrightarrow$ 3b$_2$ &
  6b$_2\longrightarrow$ 10a$_1$ &
  1b$_1 \longrightarrow$ 2a$_2$ &
  8a$_1 \longrightarrow$ 8b$_2$ \\ \hline
EE &
  LR-pCCD+S (pCCD) &
  0.3310 &
  0.3930 &
  0.4457 &
  0.4992 &
  0.5113 &
  0.5391 \\
 &
  LR-pCCD+S (HF) &
  0.3228 &
  0.3926 &
  0.4428 &
  0.5095 &
  0.5127 &
  0.5355 \\
 &
  LR-CCSD &
  0.3311 &
  0.3870 &
  0.4356 &
  0.4957 &
  0.5106 &
  0.5304 \\
 &
  EOM-CCSD &
  0.3311 &
  0.3870 &
  0.4356 &
  0.4957 &
  0.5106 &
  0.5304 \\
 &
  CCSDR(3) &
  0.3281 &
  0.3820 &
  0.4176 &
  0.4910 &
  0.4958 &
  0.5230 \\ \hline
TDM &
  LR-pCCD+S (pCCD) &
  0.3681 &
  0.1934 &
  0.7434 &
  0.9985 &
  1.0096 &
  0.7622 \\
 &
  LR-pCCD+S (HF) &
  0.2963 &
  0.1054 &
  0.8498 &
  0.9732 &
  0.9244 &
  0.9208 \\
 &
  LR-CCSD &
  0.6143 &
  0.0200 &
  0.6806 &
  1.1561 &
  1.1317 &
  0.9117 \\
 &
  EOM-CCSD &
  0.6158 &
  0.0000 &
  0.6752 &
  1.1574 &
  1.1436 &
  0.9123 \\ \hline
DS &
  LR-pCCD+S (pCCD) &
  0.1355 &
  0.0374 &
  0.5526 &
  0.9970 &
  1.0192 &
  0.5810 \\
 &
  LR-pCCD+S (HF) &
  0.0878 &
  0.0111 &
  0.7222 &
  0.9471 &
  0.8546 &
  0.8479 \\
 &
  LR-CCSD &
  0.3774 &
  0.0004 &
  0.4632 &
  1.3366 &
  1.2808 &
  0.8312 \\
 &
  EOM-CCSD &
  0.3792 &
  0.0000 &
  0.4559 &
  1.3396 &
  1.3079 &
  0.8323 \\ \hline
OS &
  LR-pCCD+S (PCCD) &
  0.0299 &
  0.0098 &
  0.1642 &
  0.3318 &
  0.3474 &
  0.2088 \\
 &
  LR-pCCD+S (HF) &
  0.0189 &
  0.0029 &
  0.2132 &
  0.3217 &
  0.2921 &
  0.3027 \\
 &
  LR-CCSD &
  0.0833 &
  0.0001 &
  0.1345 &
  0.4417 &
  0.4360 &
  0.2939 \\
 &
  EOM-CCSD &
  0.0837 &
  0.0000 &
  0.1324 &
  0.4427 &
  0.4452 &
  0.2943 \\ \hline
 &
   &
  \multicolumn{6}{c}{cc-pVTZ} \\ \cline{3-8} 
 &
   &
  1a$_2 \longrightarrow$ 7b$_2$ &
  8a$_1 \longrightarrow$ 3b$_2$ &
  1b$_1 \longrightarrow$ 3b$_2$ &
  6b$_2\longrightarrow$ 10a$_1$ &
  1b$_1 \longrightarrow$ 2a$_2$ &
  8a$_1 \longrightarrow$ 8b$_2$ \\ \hline
EE  & LR-pCCD+S (pCCD) & 0.3381 & 0.3653 & 0.4346 & 0.4773 & 0.4987 & 0.5025 \\
    & LR-pCCD+S (HF)   & 0.3515 & 0.3715 & 0.4318 & 0.4783 & 0.4836 & 0.5363 \\
    & LR-CCSD          & 0.3073 & 0.3647 & 0.4301 & 0.4717 & 0.4872 & 0.5027 \\
    & EOM-CCSD         & 0.3073 & 0.3647 & 0.4302 & 0.4717 & 0.4873 & 0.5030 \\
    & CCSDR(3)         & 0.3044 & 0.3605 & 0.4125 & 0.4671 & 0.4806 & 0.4883 \\ \hline
TDM & LR-pCCD+S (pCCD) & 0.3599 & 0.3305 & 0.6836 & 0.4062 & 0.3077 & 0.3554 \\
    & LR-pCCD+S (HF)   & 0.3432 & 0.2748 & 0.6785 & 0.6368 & 0.22   & 0.2035 \\
    & LR-CCSD          & 0.783  & 0.1892 & 0.6402 & 0.9522 & 0.2707 & 0.2406 \\
    & EOM-CCSD         & 0.7852 & 0.1913 & 0.6344 & 0.9557 & 0.2694 & 0.2478 \\ \hline
DS  & LR-pCCD+S (pCCD) & 0.1295 & 0.1092 & 0.4673 & 0.1650 & 0.0947 & 0.1263 \\
    & LR-pCCD+S (HF)   & 0.1178 & 0.0755 & 0.4603 & 0.4055 & 0.0484 & 0.0414 \\
    & LR-CCSD          & 0.6131 & 0.0358 & 0.4098 & 0.9066 & 0.0733 & 0.0579 \\
    & EOM-CCSD         & 0.6165 & 0.0366 & 0.4024 & 0.9133 & 0.0726 & 0.0614 \\ \hline
OS  & LR-pCCD+S (PCCD) & 0.0292 & 0.0266 & 0.1354 & 0.0525 & 0.0315 & 0.0423 \\
    & LR-pCCD+S (HF)   & 0.0276 & 0.0187 & 0.1325 & 0.1293 & 0.0156 & 0.0148 \\
    & LR-CCSD          & 0.1256 & 0.0087 & 0.1175 & 0.2851 & 0.0238 & 0.0194 \\
    & EOM-CCSD         & 0.1263 & 0.0089 & 0.1154 & 0.2872 & 0.0236 & 0.0206 \\ \hline
 &
   &
  \multicolumn{6}{c}{cc-pVQZ} \\ \cline{3-8} 
 &
   &
  1a$_2 \longrightarrow$ 7b$_2$ &
  8a$_1 \longrightarrow$ 3b$_2$ &
  1b$_1 \longrightarrow$ 3b$_2$ &
  6b$_2\longrightarrow$ 10a$_1$ &
  1b$_1 \longrightarrow$ 2a$_2$ &
  8a$_1 \longrightarrow$ 8b$_2$ \\ \hline
EE  & LR-pCCD+S (pCCD) & 0.3189 & 0.3653 & 0.4432 & 0.4822 & 0.5035 & 0.5189 \\
    & LR-pCCD+S (HF)   & 0.3228 & 0.3750 & 0.4307 & 0.4740 & 0.4958 & 0.5213 \\
    & LR-CCSD          & 0.2910 & 0.3630 & 0.4286 & 0.4461 & 0.4669 & 0.4819 \\
    & EOM-CCSD         & 0.2910 & 0.3630 & 0.4286 & 0.4461 & 0.4669 & 0.4819 \\
    & CCSDR(3)         & 0.3133 & 0.3515 & 0.4215 & 0.4415 & 0.4772 & 0.4910 \\ \hline
TDM & LR-pCCD+S (pCCD) & 0.4682 & 0.3033 & 0.4207 & 1.2361 & 0.9446 & 0.5384 \\
    & LR-pCCD+S (HF)   & 0.4583 & 0.3453 & 0.3451 & 1.1902 & 0.8126 & 0.3701 \\
    & LR-CCSD          & 0.1572 & 0.1364 & 0.6179 & 1.3460  & 1.1515 & 0.1476 \\
    & EOM-CCSD         & 0.1622 & 0.1393 & 0.5981 & 1.3461 & 1.1519 & 0.1487 \\ \hline
DS  & LR-pCCD+S (pCCD) & 0.2192 & 0.0920  & 0.1770 & 1.5280  & 0.8923 & 0.2899 \\
    & LR-pCCD+S (HF)   & 0.2100   & 0.1192 & 0.1191 & 1.4165 & 0.6604 & 0.1370  \\
    & LR-CCSD          & 0.0247 & 0.0186 & 0.3818 & 1.8117 & 1.3259 & 0.0218 \\
    & EOM-CCSD         & 0.0263 & 0.0194 & 0.3577 & 1.8120  & 1.3268 & 0.0221 \\ \hline
OS  & LR-pCCD+S (PCCD) & 0.0466 & 0.0224 & 0.0523 & 0.4912 & 0.2995 & 0.1003 \\
    & LR-pCCD+S (HF)   & 0.0452 & 0.0298 & 0.0342 & 0.4476 & 0.2183 & 0.0476 \\
    & LR-CCSD          & 0.0048 & 0.0045 & 0.1091 & 0.5388 & 0.4127 & 0.0070 \\
    & EOM-CCSD         & 0.0051 & 0.0047 & 0.1022 & 0.5389 & 0.4130 & 0.0071\\ \hline
\end{tabular}
\end{table*}
\fancypagestyle{plain}{
\fancyhf{}
\renewcommand{\headrulewidth}{as_u_prefer_here} 
\rhead{\textbf{Table S4 (continued)}}
}
\FloatBarrier
\clearpage
\thispagestyle{mypagestyle3}  

\begin{table}
\centering
\renewcommand{\arraystretch}{0.83}
\setlength\tabcolsep{6pt} 
\DefTblrTemplate{firsthead, middlehead,lasthead}{default}{}
\begin{tabular}{llcccccc}
 &
  \multicolumn{6}{c}{aug-cc-pVTZ} \\ \hline \\
 Property & Method&
  1a$_2 \longrightarrow$ 7b$_2$ &
  8a$_1 \longrightarrow$ 3b$_2$ &
  1b$_1 \longrightarrow$ 3b$_2$ &
  6b$_2\longrightarrow$ 10a$_1$ &
  1b$_1 \longrightarrow$ 2a$_2$ &
  8a$_1 \longrightarrow$ 8b$_2$ \\ \hline \\
EE  & LR-pCCD+S (pCCD) & 0.3189 & 0.4432 & 0.3653 & 0.4822 & 0.5035 & 0.5189 \\
    & LR-pCCD+S (HF)   & 0.2434 & 0.3268 & 0.3198 & 0.3307 & 0.4059 & 0.4097 \\
    & LR-CCSD          & 0.2465 & 0.3225 & 0.3079 & 0.3306 & 0.4007 & 0.4097 \\
    & EOM-CCSD         & 0.2465 & 0.3225 & 0.3079 & 0.3306 & 0.4007 & 0.4097 \\
    & CCSDR(3)         & 0.2444 & 0.3198 & 0.3016 & 0.3264 & 0.3972 & 0.4077 \\ \\ \hline\\
TDM & LR-pCCD+S (pCCD) & 0.3904 & 0.3971 & 0.4175 & 0.5066 & 0.3947 & 0.3697 \\
    & LR-pCCD+S (HF)   & 0.4717 & 0.4988 & 0.8843 & 0.7793 & 0.4610 & 0.3138 \\
    & LR-CCSD          & 0.4909 & 0.2390 & 1.3828 & 0.8347 & 0.2733 & 0.1806 \\
    & EOM-CCSD         & 0.4934 & 0.2415 & 1.4179 & 0.8594 & 0.2722 & 0.1758 \\ \\ \hline\\
DS  & LR-pCCD+S (pCCD) & 0.1524 & 0.1577 & 0.1743 & 0.2566 & 0.1558 & 0.1367 \\
    & LR-pCCD+S (HF)   & 0.2225 & 0.2488 & 0.7819 & 0.6073 & 0.2125 & 0.0985 \\
    & LR-CCSD          & 0.2410 & 0.0571 & 1.9120 & 0.6968 & 0.0747 & 0.0326 \\
    & EOM-CCSD         & 0.2434 & 0.0583 & 2.0103 & 0.7385 & 0.0741 & 0.0309 \\ \\ \hline\\
OS  & LR-pCCD+S (PCCD) & 0.0324 & 0.0466 & 0.0515 & 0.0825 & 0.0523 & 0.0473 \\
    & LR-pCCD+S (HF)   & 0.0361 & 0.0542 & 0.1667 & 0.1339 & 0.0575 & 0.0269 \\
    & LR-CCSD          & 0.0396 & 0.0123 & 0.3925 & 0.1536 & 0.0199 & 0.0089 \\
    & EOM-CCSD         & 0.0400 & 0.0125 & 0.4127 & 0.1628 & 0.0198 & 0.0085 \\ \hline\\
\end{tabular}
\end{table}
\clearpage
\begin{figure*}
    \centering
    \includegraphics[width=\textwidth]{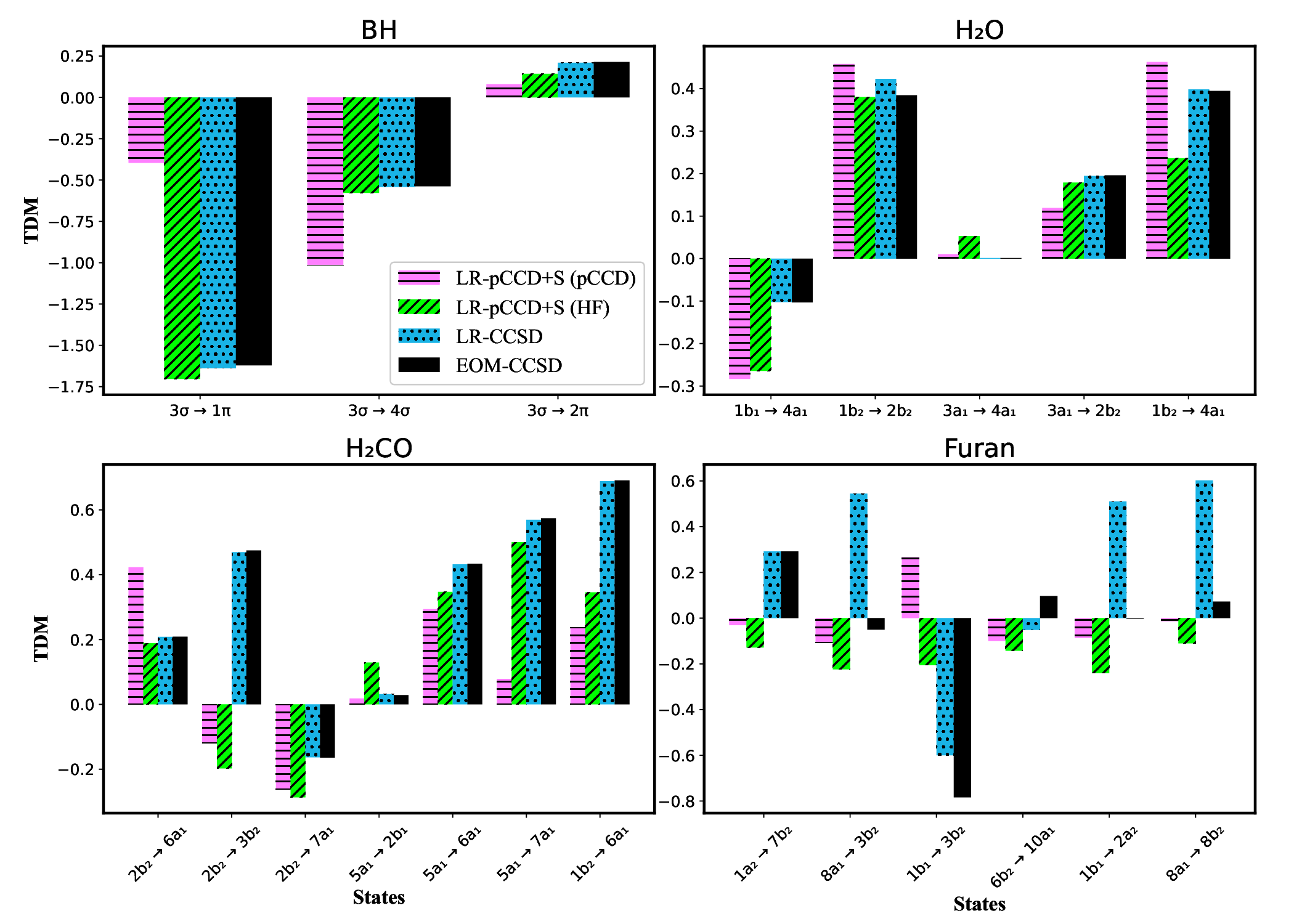}
    \caption{Difference in TDM between the cc-pVTZ and aug-cc-pVTZ basis sets for BH, \ce{H2O}, \ce{H2CO}, and Furan molecules.}
    \label{fig:tz-atz}
\end{figure*}
\begin{table*}
\begin{center}
\caption{Mean Absolute Error (MA), Mean Average Error (ME), and Standard Deviation (SD) for the LR-pCCD+S TDM, OS, and DS of BH, \ce{H2O}, \ce{H2CO}, and Furan molecules w.r.t. LR-CCSD. dz, tz, and qz denote cc-pVDZ, cc-pVTZ, and cc-pVQZ basis sets, respectively. }\label{tab:error}
\begin{tabular}{llccc|ccc}
\multicolumn{5}{c|}{BH}                                      & \multicolumn{3}{c}{\ce{H2O}}   \\ \hline
Property             & Basis      & dz   & tz   & qz   & dz    & tz   & qz   \\ \hline
\multirow{4}{*}{TDM} & SD (pCCD)$ ^a$  & 0.0482 & 0.0727 & 0.0285 & 0.0081 & 0.0101 & 0.0198 \\
                     & SD (HF)$ ^b$    & 0.0474 & 0.0250 & 0.0140 & 0.0113 & 0.0062 & 0.0132 \\
                     & MAE (pCCD) $ ^c$ & 0.0878 & 0.1117 & 0.0567 & 0.0179 & 0.0205 & 0.0363 \\
                     & MAE (HF)$ ^d$   & 0.0931 & 0.0199 & 0.0361 & 0.0234 & 0.0158 & 0.0253 \\ \hline
\multirow{4}{*}{DS}  & SD (pCCD)  & 0.1282 & 0.1625 & 0.0355 & 0.0106 & 0.0115 & 0.0252 \\
                     & SD (HF)    & 0.1260 & 0.0484 & 0.0303 & 0.0143 & 0.0091 & 0.0169 \\
                     & MAE (pCCD) & 0.2304 & 0.2294 & 0.0810 & 0.0219 & 0.0243 & 0.0444 \\
                     & MAE (HF)   & 0.2403 & 0.0494 & 0.0618 & 0.0287 & 0.0200 & 0.0304 \\
\multirow{4}{*}{OS}  & SD (pCCD)  & 0.0299 & 0.0326 & 0.0069 & 0.0041 & 0.0048 & 0.0103 \\ \hline
                     & SD (HF)    & 0.0301 & 0.0107 & 0.0049 & 0.0181 & 0.0032 & 0.0049 \\
                     & MAE (pCCD) & 0.0550 & 0.0444 & 0.0158 & 0.0330 & 0.0090 & 0.0185 \\
                     & MAE (HF)   & 0.0538 & 0.0083 & 0.0115 & 0.0323 & 0.0070 & 0.0118\\ \hline
                     & \multicolumn{4}{c|}{\ce{H2CO}}             & \multicolumn{3}{c}{Furan} \\ \hline
\multirow{4}{*}{TDM} & SD (pCCD)  & 0.0192 & 0.0139 & 0.0076 & 0.0108 & 0.0227 & 0.0369 \\
                     & SD (HF)    & 0.0131 & 0.0041 & 0.0024 & 0.0143 & 0.0255 & 0.0112 \\
                     & MAE (pCCD) & 0.0299 & 0.0268 & 0.0118 & 0.0264 & 0.0708 & 0.0536 \\
                     & MAE (HF)   & 0.0220 & 0.0147 & 0.0061 & 0.0281 & 0.0736 & 0.0340 \\ \hline
\multirow{4}{*}{DS}  & SD (pCCD)  & 0.0347 & 0.0234 & 0.0137 & 0.0147 & 0.0264 & 0.0588 \\
                     & SD (HF)    & 0.0219 & 0.0076 & 0.0041 & 0.0223 & 0.0284 & 0.0169 \\
                     & MAE (pCCD) & 0.0477 & 0.0385 & 0.0166 & 0.0339 & 0.0811 & 0.0750 \\
                     & MAE (HF)   & 0.0351 & 0.0221 & 0.0076 & 0.0387 & 0.0830 & 0.0494 \\ \hline
\multirow{4}{*}{OS}  & SD (pCCD)  & 0.0127 & 0.0070 & 0.0039 & 0.0049 & 0.0053 & 0.0113 \\
                     & SD (HF)    & 0.0087 & 0.0018 & 0.0017 & 0.0069 & 0.0056 & 0.0034 \\
                     & MAE (pCCD) & 0.0164 & 0.0124 & 0.0048 & 0.0108 & 0.0162 & 0.0154 \\
                     & MAE (HF)   & 0.0102 & 0.0066 & 0.0012 & 0.0120 & 0.0164 & 0.0105 \\ \hline
\end{tabular}
\end{center}
\begin{small}
\vspace{0.1cm}
\text{\hspace{1.1cm}\( ^a \hspace{0.2cm} \textrm{SD} = \sqrt{\frac{\sum_n(x_n-\bar{x})^2}{N}}\), and SD(pCCD) is DS for pCCD+S(pCCD) results, $\bar{x}$ is average over all states and $N$ is number of states.}\\
\text{\hspace{1cm}\( ^b \hspace{0.2cm}\) SD(HF) is DS for pCCD+S(HF) results.}

\text{\( \hspace{1.4cm} \textrm{MAE} = \frac{1}{N} \sum_n |x_n-x_r|\), where $x_r$ is a LR-CCSD results as a reference data.}

\text{\hspace{1cm}\( ^c\hspace{0.2cm}\) MAE(pCCD) is MAE, where \(x_a = \)LR-pCCD+S(pCCD)results.}

\text{\hspace{1cm}\( ^d \hspace{0.2cm} \) MAE(HF) is MAE, where \(x_a = \)LR-pCCD+S(HF) results.}
\end{small}
\end{table*}
\begin{table*}
\caption{Average of standard deviations (SDs) and mean absolute errors (MAEs) of data in Table~\ref{tab:error}, over all molecules. }
\begin{center}
\begin{tabular}{lllll}
Property             & Statistics & cc-pVDZ & cc-pVTZ & cc-pVQZ \\ \hline
\multirow{4}{*}{TDM} & SD (pCCD)  & 0.0216 & 0.0299 & 0.0232 \\
                     & SD (HF)    & 0.0215 & 0.0152 & 0.0102 \\
                     & MAE (pCCD) & 0.0405 & 0.0575 & 0.0396 \\
                     & MAE (HF)   & 0.0416 & 0.0310 & 0.0254 \\
\multirow{4}{*}{DS}  & SD (pCCD)  & 0.0470 & 0.0559 & 0.0333 \\
                     & SD (HF)    & 0.0461 & 0.0234 & 0.0171 \\
                     & MAE (pCCD) & 0.0835 & 0.0933 & 0.0542 \\
                     & MAE (HF)   & 0.0857 & 0.0436 & 0.0373 \\
\multirow{4}{*}{OS}  & SD (pCCD)  & 0.0129 & 0.0124 & 0.0081 \\
                     & SD (HF)    & 0.0160 & 0.0053 & 0.0037 \\
                     & MAE (pCCD) & 0.0288 & 0.0205 & 0.0136 \\
                     & MAE (HF)   & 0.0271 & 0.0096 & 0.0087 \\ \hline
\end{tabular}
\end{center}
\end{table*}
\clearpage
\section{Deriving the Jacobian matrix elements}
To obtain the linear response vectors and matrices, we consider the de-excitation, and excitation operators, summarized in Table~1. The molecular Hamiltonian is defined as
    \begin{align}
           \mathbf{H}_0 = \sum_{pq} f_p^q \{\hat{p}^{\dagger} \hat{q}\} + \frac{1}{2}\sum_{pqrs} V_{pq}^{rs} \{\hat{p}^{\dagger} \hat{q}^{\dagger} \hat{s} \hat{r}\},
    \end{align}
and the second quantization operator commutation relations are,
    \begin{align}
          &[\hat{p}^{\dagger},\hat{P}_b]=-\hat{b} \, \delta_{pb}\,, \hspace{0.5cm}[\hat{r},\hat{P}_i^{\dagger}]=\hat{i}^{\dagger}\, \delta_{ri},\nonumber \\
          & [\hat{i}^{\dagger},\hat{P}_b]= [\hat{i},\hat{P}_{a}^{\dagger}]=[\hat{a}^{\dagger},\hat{P}_i]= [\hat{a},\hat{P}_i^{\dagger}]= 0.
    \end{align}

\subsection{The Jacobian matrix for pair excitations}

The Jacobian matrix elements for the pair excitations are as follow
    \begin{align}
            \boldsymbol{J}_{\mu_2\nu_2} &=\boldsymbol{J}_{iajb}^{(2,2)}= \langle \mu_2| [\bar{{H}}_0,\hat{\tau}_{\nu_2}]|\textrm{HF}\rangle
            = \langle \textrm{HF}|\hat{\tau}_{\mu_2}^{\dagger}[\hat{H}_0,\hat{\tau}_{\nu_2}]|\textrm{HF}\rangle \nonumber\\
            &  + \sum_{kc}\langle \textrm{HF}|\hat{\tau}_{\mu_2}^{\dagger} [[\hat{H}_0,t_{k\Bar{k}}^{c\Bar{c}}\hat{\tau}_{k_2}],\hat{\tau}_{\nu_2}]|\textrm{HF}\rangle
            \nonumber\\
            &= \langle \textrm{HF}| \hat{P}_i^{\dagger} \hat{P}_a  [\hat{H}_0, \hat{P}_b^{\dagger} \hat{P}_j]|\textrm{HF}\rangle \nonumber\\
            &  + \langle \textrm{HF}| \hat{P}_i^{\dagger} \hat{P}_a  [[\hat{H}_0 ,\sum_{ck} \cia{k}{c} 
           \hat{P}_c^{\dagger} \hat{P}_k] , \hat{P}_b^{\dagger} \hat{P}_j]|\textrm{HF}\rangle.
    \end{align}

For pair-pair couplings,
    \begin{align}
          J_{iajb}^{(2,2)} &= \langle \textrm{HF}|\{\hat{i}^{\dagger}\hat{\Bar{i}}^{\dagger} \hat{\Bar{a}}\hat{a} \} \big(\sum_{pq} f_{p}^{q} \{\hat{p}^{\dagger}\hat{q}\} +\frac{1}{4}\sum_{pqrs} V_{pq}^{rs} \{\hat{p}^{\dagger}\hat{q}^{\dagger}\hat{s} \hat{r}\}\big) \{\hat{b}^{\dagger}\hat{\Bar{b}}^{\dagger} \hat{\Bar{j}}\hat{j} \}|\textrm{HF}\rangle \nonumber\\
           &  +\langle \textrm{HF}|\{\hat{i}^{\dagger}\hat{\Bar{i}}^{\dagger} \Bar{a}a \} \big( \sum_{pq}f_{p}^{q} \{\hat{p}^{\dagger}\hat{q}\} + \frac{1}{4}\sum_{pqrs} V_{pq}^{rs} \{\hat{p}^{\dagger}\hat{q}^{\dagger}\hat{s} \hat{r}\}\big) \nonumber \\
           &   \sum_{kc} \cia{k}{c} \{c^{\dagger}\Bar{c}^{\dagger} \Bar{k}k \}  \{\hat{b}^{\dagger}\hat{\Bar{b}}^{\dagger} \hat{\Bar{j}}\hat{j} \}|\textrm{HF}\rangle. 
    \end{align}

Using Wick's theorem to identify the connected terms, the pCCD Jacobian matrix can be written as
    \begin{align}
            \boldsymbol{J}_{iajb}^{(2,2)} &= \big( \,2f_{a}^{a} -2 f_{i}^{i} + 2 V_{ia}^{ai} - 4 V_{ia}^{ia} -2\sum_{k}V_{kk}^{bb} \cia{k}{a} \nonumber\\
           &- 2\sum_{c}V_{j}^{c} \cia{i}{c} + 4 V_{jj}^{bb} \cia{i}{a} \big)\,\delta_{ab}\,\delta_{ij} \nonumber\\
           &  +\big( V_{ii}^{jj}-2V_{jj}^{bb} \cia{i}{a} + \sum_{c}V_{jj}^{cc} \cia{i}{c} \big)\, \delta_{ab} \nonumber\\
           &  +\big( V_{bb}^{aa}-2V_{jj}^{bb} \cia{i}{a} + \sum_{k}V_{kk}^{bb} \cia{k}{a} \big)\, \delta_{ij}.
    \end{align}

\subsection{The Jacobian matrix for single excitations}

Similarly, for single-single couplings,
    \begin{align}
            \boldsymbol{J}_{\mu_1\nu_1} &=\boldsymbol{J}_{iajb}^{(1,1)}= \langle \mu_1| [\bar{{H}}_0,\hat{\tau}_{\nu_1}]|\textrm{HF} \rangle
            = \langle \textrm{HF}|\hat{\tau}_{\mu_1}^{\dagger}[\hat{H}_0,\hat{\tau}_{\nu_1}]| \textrm{HF}\rangle \nonumber\\
            &  + \sum_{kc}\langle \textrm{HF}|\hat{\tau}_{\mu_1}^{\dagger} [[\hat{H}_0,t_{k\Bar{k}}^{c\Bar{c}}\hat{\tau}_{k_2}],\hat{\tau}_{\nu_1}]|\textrm{HF} \rangle
            \nonumber\\
            &= \langle \textrm{HF}| \hat{i}^{\dagger} \hat{a}  [\hat{H}_0, \hat{b}^{\dagger} \hat{j}]|\textrm{HF}\rangle \nonumber\\
            &  + \langle \textrm{HF}| \hat{i}^{\dagger} \hat{a}  [[\hat{H}_0 ,\sum_{ck} \cia{k}{c} 
           \hat{P}_c^{\dagger} \hat{P}_k] , \hat{b}^{\dagger} \hat{j}]|\textrm{HF}\rangle,
    \end{align}

    \begin{align}
            \boldsymbol{J}_{iajb}^{(1,1)} &= (2f_{a}^{b} -2\sum_k \cia{k}{c} V_{kk}^{ab})\,\delta_{ij} \nonumber \\
            &  -2f_{j}^{i} -2\sum_c \cia{i}{c} V_{ij}^{cc})\,\delta_{ab} \nonumber \\ 
            &  + 4V_{ja}^{bi} - 2V_{aj}^{bi} + \cia{i}{a}(4V_{ij}^{ab} - 2V_{ij}^{ba}).
    \end{align}

\subsection{The Jacobian matrix for pair-single excitations}

For pair-single couplings,
    \begin{align}
            \boldsymbol{J}_{\mu_2\nu_1} &=\boldsymbol{J}_{iajb}^{(2,1)}= \langle \mu_2| [\bar{{H}}_0,\hat{\tau}_{\nu_1}]| \textrm{HF} \rangle
            = \langle \textrm{HF}|\hat{\tau}_{\mu_2}^{\dagger}[\hat{H}_0,\hat{\tau}_{\nu_1}]|\textrm{HF}\rangle \nonumber\\
            &  + \sum_{kc}\langle \textrm{HF}|\hat{\tau}_{\mu_2}^{\dagger} [[\hat{H}_0,t_{k\Bar{k}}^{c\Bar{c}}\hat{\tau}_{k_2}],\hat{\tau}_{\nu_1}]|\textrm{HF}\rangle
            \nonumber\\
            &= \langle \textrm{HF}| \hat{P}_i^{\dagger} \hat{P}_a  [\hat{H}_0, \hat{P}_b^{\dagger} \hat{P}_j]|\textrm{HF}\rangle \nonumber\\
            &  + \langle \textrm{HF}| \hat{P}_i^{\dagger} \hat{P}_a  [[\hat{H}_0 ,\sum_{ck} \cia{k}{c} 
           \hat{P}_c^{\dagger} \hat{P}_k] , \hat{b}^{\dagger} \hat{j}]|\textrm{HF}\rangle.
    \end{align}

    \begin{align}
            \boldsymbol{J}_{iajb}^{(2,1)} &= -2 \cia{i}{a} f_{j}^{a} + \cia{i}{a} (4V_{ij}^{ia} - 2V_{ij}^{ai}) - 4 \sum_c \cia{i}{c} V_{ja}^{cc} \,\delta_{ab} \nonumber \\ 
            &  -2 \cia{i}{a} f_{i}^{b} - \cia{i}{a} (4V_{ai}^{ab} - 2V_{ia}^{ab}) - 4 \sum_k \cia{k}{a} V_{kk}^{bi} \,\delta_{ij} \nonumber \\
            &  - \cia{i}{a} (4(V_{aj}^{ab} - V_{ji}^{bi}) - 2(V_{ja}^{ai} - V_{ai}^{ba})).
    \end{align}

\subsection{The Jacobian matrix for single-pair excitations}

Finally, for single-pair couplings,
    \begin{align}
            \boldsymbol{J}_{\mu_1\nu_2} &=\boldsymbol{J}_{iajb}^{(1,2)}= \langle \mu_1| [\bar{{H}}_0,\hat{\tau}_{\nu_2}]|\textrm{HF} \rangle
            = \langle \textrm{HF}|\hat{\tau}_{\mu_1}^{\dagger}[\hat{H}_0,\hat{\tau}_{\nu_2}]|\textrm{HF} \rangle \nonumber\\
            &  + \sum_{kc}\langle \textrm{HF}|\hat{\tau}_{\mu_1}^{\dagger} [[\hat{H}_0,t_{k\Bar{k}}^{c\Bar{c}}\hat{\tau}_{k_2}],\hat{\tau}_{\nu_2}]|\textrm{HF}\rangle
            \nonumber\\
            &= \langle \textrm{HF}| \hat{i}^{\dagger} \hat{a}  [\hat{H}_0, \hat{P}_b^{\dagger} \hat{P}_j]|\textrm{HF} \rangle \nonumber\\
            &  + \langle \textrm{HF}| \hat{i}^{\dagger} \hat{a}  [[\hat{H}_0 ,\sum_{ck} \cia{k}{c} 
           \hat{P}_c^{\dagger} \hat{P}_k] , \hat{P}_b^{\dagger} \hat{P}_j]|\textrm{HF} \rangle.
    \end{align}

    \begin{align}
            \boldsymbol{J}_{iajb}^{(1,2)} &= -2 \cia{i}{a} f_{j}^{a} - \cia{i}{a} (4V_{ij}^{ia} - 2V_{ij}^{ai}) - 4 \sum_c \cia{i}{c} V_{ja}^{cc} \,\delta_{ab} \nonumber \\ 
            &  -2 \cia{i}{a} f_{i}^{b} + \cia{i}{a} (4V_{ai}^{ab} - 2V_{ia}^{ab}) - 4 \sum_k \cia{k}{a} V_{kk}^{bi} \,\delta_{ij} \nonumber \\
            &  - \cia{i}{a} (4(V_{aj}^{ab} - V_{ji}^{bi}) - 2(V_{ja}^{ai} - V_{ai}^{ba})).
    \end{align}

The elements $J_{\mu_2\nu_2}$, $J_{\mu_1\nu_1}$, $J_{\mu_2\nu_1}$, and $J_{\mu_1\nu_2}$ correspond to different excitation types. Each matrix element is derived from the expectation values of the Hamiltonian operator, appropriately expanded with Wick's theorem and commutators. For detailed derivations, refer to the primary coupled-cluster theory literature.

This formalism outlines the construction of the pCCD Jacobian matrix, providing a foundation for further computational implementation and analysis of electron correlation effects in molecular systems.
The final Jacobian matrix can be written as,
    \begin{align}\label{eqn:jacobian}
            J_{iajb} = \begin{pmatrix}
                    J_{iajb}^{(1,1)} & & J_{iajb}^{(1,2)}\\
                    \\
                    J_{iajb}^{(2,1)} & & J_{iajb}^{(2,2)} 
            \end{pmatrix}.
    \end{align}

\section{Derivation of the $F_{\mu\nu}$ matrix elements}
The transition matrix elements for \(F_{\mu\nu}\) (see also Table~1) can be derived using the following equations,
    \begin{align}
        F_ {\nu k} &=  \langle \Lambda|[[\hat{H}_0,\hat{\tau}_\nu], \hat{\tau}_k]|\textrm{pCCD}\rangle,
    \end{align}
for pair-pair couplings,
    \begin{align}
          F_{\mu_2 \nu_2} &= F_{iajb}^{(2,2)} =  \langle \Lambda [[\hat{H}_0,\hat{\tau}_{\mu_2}], \hat{\tau}_{\nu_2}]|\textrm{pCCD}\rangle \nonumber \\
           &= (\cia{i}{b}V_{jj}^{aa} + \cia{j}{a}V_{ii}^{bb}) + 4(\lia{i}{a}V_{ii}^{aa})\delta_{ij}\delta_{ab} \nonumber \\
           &  -2(\lia{i}{b}V_{jj}^{aa} + \lia{i}{a} V_{jj}^{bb})\delta_{ij} \nonumber \\
           &  -2(\lia{i}{a} V_{jj}^{bb} + \lia{j}{a} V_{ii}^{bb})\delta_{ab}.
    \end{align}
For single-single couplings,
    \begin{align}
          F_{\mu_1 \nu_1} &= F_{iajb}^{(1,1)} =  \langle \Lambda [[\hat{H}_0,\hat{\tau}_{\mu_1}], \hat{\tau}_{\nu_1}]|\textrm{pCCD}\rangle \nonumber \\
           &=-2\lia{i}{b}(2V_{jb}^{ai} - V_{jb}^{ia}) - 2\lia{j}{a} (2V_{ia}^{bj} - V_{ia}^{jb}) \nonumber \\
           &  + 2(2V_{ij}^{ab} - V_{ji}^{ab}) + (\lia{i}{b}\cia{i}{b} + \lia{j}{a} \cia{j}{a})(2V_{ji}^{ba} - V_{ji}^{ab}) \nonumber \\
           &  -2\sum_l \lia{l}{b}\cia{l}{b}(2V_{ij}^{ab} - V_{ji}^{ba}) \nonumber \\
           &  + (\sum_{lc} \lia{l}{c}\cia{l}{c} V_{ij}^{cc} + 2\sum_d \lia{i}{d} V_{dd}^{ba})\delta_{ij} \nonumber \\
           &  + (\sum_{dk} \lia{j}{d}\cia{k}{d} V_{kk}^{ba} + 2\sum_l \lia{l}{a} V_{ji}^{ll})\delta_{ab}.
    \end{align}
For single-pair couplings,
    \begin{align}
          F_{\mu_1 \nu_2} &= F_{iajb}^{(1,2)} =  \langle \Lambda [[\hat{H}_0,\hat{\tau}_{\mu_1}], \hat{\tau}_{\nu_2}]|\textrm{pCCD}\rangle \nonumber \\
           &= \lia{i}{a} (2V_{ji}^{ai} - 2V_{ij}^{ai} - 2V_{jb}^{aa} + 4V_{ja}^{ba} - 4V_{ji}^{bi}) + 4\lia{j}{a} V_{ii}^{bj} \nonumber \\
           &  -2 (\lia{i}{a} (f_{i}^{b} + 2V_{ia}^{ba}))\delta_{ij} \nonumber \\
           &  -2 \lia{i}{b}f_{j}^{b} + 2 \lia{i}{a} (2V_{ij}^{ia} - V_{ij}^{ai})\delta_{ab}.
    \end{align}
For pair-single couplings,
    \begin{align}
          F_{\mu_2 \nu_1} &= F_{iajb}^{(2,1)} =  \langle \Lambda [[\hat{H}_0,\hat{\tau}_{\mu_2}], \hat{\tau}_{\nu_1}]|\textrm{pCCD}\rangle \nonumber \\
           &= \lia{j}{b} (2V_{ij}^{bj} + 2V_{ij}^{ja} + 2V_{jj}^{ia} - 2V_{jb}^{ab} + 2V_{ib}^{ab} - 4V_{ij}^{aj}) \nonumber \\
           &  -2 ( \lia{j}{b} (f_{j}^{a} + 2V_{jb}^{ab}))\delta_{ij} \nonumber \\
           &  -2(\lia{j}{a} (f_{i}^{b} + 2(V_{ij}^{jb} - V_{ij}^{bj})))\delta_{ab}.
    \end{align}
Thus, the \(F_{iajb}\) matrix can be expressed as,
    \begin{align}
        F_{iajb} = \begin{pmatrix}
                F_{iajb}^{(1,1)} & & F_{iajb}^{(1,2)}\\
                \\
                F_{iajb}^{(2,1)} & & F_{iajb}^{(2,2)} 
        \end{pmatrix}.
    \end{align}

\section{Derivation of the $\xi_{\nu}^{\hat{A}}$ vector elements}
The vector \(\xi_{\nu}^{\hat{A}}\)(Table~1) when \(\hat{A}\) is the dipole operator can be written for pair and single parts, respectively, as the following equations,
    \begin{align}
         \xi_{\nu_2}^{\hat{D}} &= \langle\nu_2|\bar{{D}}|\textrm{HF}\rangle = \langle\nu_2|\hat{D}|\textrm{HF}\rangle + \langle\nu_2|\hat{D} \Tp |\textrm{HF}\rangle \nonumber \\
         &= 2 \cia{i}{a} (d_{aa} - d_{ii}),
    \end{align}
        
    \begin{align}
         \xi_{\nu_1}^{\hat{D}} &= \langle\nu_1|\bar{{D}}|\textrm{HF}\rangle = \langle\nu_1|\hat{D}|\textrm{HF}\rangle + \langle\nu_1|\hat{D} \Tp |\textrm{HF}\rangle \nonumber \\
         &= 2 \cia{i}{a} d_{ia}.
    \end{align}

Thus, the vector \(\xi_{\nu}^{\hat{D}}\) can be expressed as,
    \begin{align}
        \xi_{\nu}^{\hat{D}} = \begin{pmatrix}
            \xi_{\nu_1}^{\hat{D}}, & & \xi_{\nu_2}^{\hat{D}}
        \end{pmatrix}.
    \end{align}

\section{Derivation of the $\eta_{\nu}^{\hat{A}}$ vector elements}
The vector \(\eta_{\nu}^{\hat{A}}\) (cf. Table~1) when \(\hat{A}\) is the dipole operator and using \(\langle \lambda |\) from Table~1 can be written for pair and single parts, respectively, as
    \begin{align}
         \eta_{\nu_2}^{\hat{D}} &= \langle\lambda|[\hat{D},\hat{\tau}_{\nu_2}]|\textrm{pCCD}\rangle \nonumber \\
         &= \langle \textrm{HF}|[\hat{D}, \hat{\tau}_{\nu_2}]|\textrm{HF}\rangle + \langle \textrm{HF}|[\hat{D},\hat{\tau}_{\nu_2}]\Tp |\textrm{HF} \rangle \nonumber \\
         &  + \sum_{\mu} \bar{t}_{\mu}\langle \mu_2|[\hat{D}, \hat{\tau}_{\nu_2}]|\textrm{HF} \rangle + \sum_{\mu} \bar{t}_{\mu}\langle \mu_2|[[\hat{D}, \hat{\tau}_{\nu_2}],\Tp]|\textrm{HF} \rangle \nonumber \\
         &= 2 \lia{j}{b} (d_{bb} - d_{jj}),
    \end{align}
and 
    \begin{align}
         \eta_{\nu_1}^{\hat{D}} &= \langle\lambda|[\hat{D},\hat{\tau}_{\nu_1}]|\textrm{pCCD}\rangle \nonumber \\
         &= \langle \textrm{HF}|[\hat{D}, \hat{\tau}_{\nu_1}]|HF\rangle + \langle \textrm{HF}|[\hat{D},\hat{\tau}_{\nu_1}]\Tp |\textrm{HF}\rangle \nonumber \\
         &  + \sum_{\mu} \bar{t}_{\mu}\langle \mu_2|[\hat{D}, \hat{\tau}_{\nu_1}]|\textrm{HF}\rangle + \sum_{\mu} \bar{t}_{\mu}\langle \mu_2|[[\hat{D}, \hat{\tau}_{\nu_1}],\Tp]|\textrm{HF}\rangle \nonumber \\
         &= -2 d_{ai} -2 d_{ia} \left(\sum_d \cia{i}{d} \lia{i}{d} + \sum_l \cia{l}{a} \lia{l}{a} \right).
    \end{align}

Thus, the vector \(\eta_{\nu}^{\hat{D}}\) can be expressed as
    \begin{align}
        \eta_{\nu}^{\hat{D}} = \begin{pmatrix}
            \eta_{\nu_1}^{\hat{D}}, & & \eta_{\nu_2}^{\hat{D}}
        \end{pmatrix}.
    \end{align}

\section{ Transition matrix equation}
I this section we consider excitation energies and transition moments, which can be obtained based on response functions in coupled cluster theory. We first focus on the exact linear response functions. We assume \( \hat{A} \) and \( \hat{B}\) are Hermitian operators,  \( |n\rangle \) and \( E_n\) are defined as exact solutions to the eigenvalue problem of the (time-independent) electronic Hamiltonian \( \hat{H} \) as follows,
    \begin{align}
        \hat{H} |n\rangle = E_n |n\rangle.
    \end{align}
We number the eigenfunctions and eigenvalues with  \( n \) (\( n = 0, 1, 2, \dots \)), where \( n = 0 \) corresponds to the ground state. We also introduce,
    \begin{align}
        \omega_n = E_n - E_0,
    \end{align}
as the excitation energy from the ground state to the excited state \( n \). The transition moment, which defines the probability of transition from state \( k \) to state \( n \) due to perturbation \( \hat{B} \), is given for exact wave functions by the expression
    \begin{align}
        T^{\hat{B}}_{kn} = \langle k | \hat{B} | n \rangle.
    \end{align}
It follows from this that for exact wave functions,
    \begin{align}
        T^{\hat{B}}_{kn} = (T^{\hat{B}}_{nk})^{\ast}.
    \end{align}
The linear response function can be represented as a sum over states,
    \begin{align}
        \langle \langle \hat{A}; \hat{B}\rangle \rangle_{\omega_{\gamma}} = P_{\hat{A} \hat{B}} \sum_{n > 0} \frac{\langle 0 | \hat{A} | n \rangle \langle n | \hat{B} |0 \rangle}{\omega_{\hat{B}} - \omega_n},
    \end{align}
where, \(\gamma=\hat{A}, \hat{B}\) and \( P_{\hat{A} \hat{B}} \) is the sum of permutations of the pairs \( ( \hat{A}, \omega_{\hat{A}} ), ( \hat{B}, \omega_{\hat{B}} ) \), remembering that \( \omega_{\hat{A}} + \omega_{\hat{B}} = 0 \). The above formula can be written as,
    \begin{align} 
        \langle \langle \hat{A}; \hat{B}\rangle \rangle_{\omega} &= \sum_{n > 0} \frac{\langle0 | \hat{A} | n \rangle \langle n | \hat{B} |0 \rangle}{\omega_{\hat{B}} - \omega_n} + \sum_{n > 0} \frac{\langle0 | \hat{B} | n \rangle \langle n | \hat{A} |0 \rangle}{\omega_{\hat{A}} - \omega_n},\\ \nonumber
        &= \sum_{n > 0} \frac{\langle 0 | \hat{A} | n \rangle \langle n | \hat{B} |0 \rangle}{\omega_{\hat{B}} - \omega_n} - \sum_{n > 0} \frac{\langle 0 | \hat{B} | n \rangle \langle n | \hat{A} |0 \rangle}{\omega_{\hat{B}} + \omega_n}.
    \end{align}
From now on, we assume that the frequencies \( \omega \) are real. The linear response function \( \langle \langle \hat{A}; \hat{B}\rangle \rangle_{\omega} \) is a function of frequency \( \omega \). As a function of frequency, it has first-order poles for values of \( \omega \) equal to the excitation energy \( +\omega_n \) and de-excitation energy \( -\omega_n \) (at these points, some denominators in the sum-over-states expansion above become zero). From the residuals of the function \( \langle \langle \hat{A}; \hat{B}\rangle \rangle_{\omega} \), we can obtain transition moments as follows,
    \begin{align} 
        &\lim_{\omega \to \omega_k} (\omega - \omega_k) \langle \langle \hat{A}; \hat{B}\rangle \rangle_{\omega} =\\ \nonumber
         \sum_{n > 0} &\lim_{\omega \to \omega_k} \frac{\omega - \omega_k}{\omega - \omega_n} \langle0 | \hat{A} | n \rangle \langle n | \hat{B} |0 \rangle -\\ \nonumber
        \sum_{n > 0} &\lim_{\omega \to \omega_k} \frac{\omega - \omega_k}{\omega + \omega_n} \langle 0 | \hat{B} | n \rangle \langle n | \hat{A} |0 \rangle = \langle 0 | \hat{A} |k \rangle \langle k | \hat{B} |0 \rangle
    \end{align}
This passage to the limit as \( \omega \to \omega_k \) omits all terms of the expansion except for the first in the first sum, where \( n = k \). Hence,
    \begin{align}
        \frac{\omega - \omega_k}{\omega - \omega_n} = 1.
    \end{align}
The obtained term is element of the transition matrix,
    \begin{align}
        \Gamma^{\hat{A} \hat{B}}_{0k} = \lim_{\omega \to \omega_k} (\omega - \omega_k) \langle \langle \hat{A}; \hat{B}\rangle \rangle_{\omega} = \underbrace{\langle 0 | \hat{A} |k \rangle}_{T^{\hat{A}}_{0k}} \underbrace{\langle k | \hat{B} |0 \rangle}_{T^{\hat{B}}_{k0}},
    \end{align}
where \( T_{0k} \) ( \( T_{k0} \)) is known as the right (left) transition matrix.

\bibliography{ArXiv_v/main_a}